\documentclass[11pt,a4paper]{article}
\pdfoutput=1

\usepackage{epsfig}
\usepackage{latexsym}
\usepackage{amsfonts}
\usepackage{amsmath}
\usepackage{amsthm}
\usepackage{amssymb}
\usepackage{amsbsy}
\usepackage{multirow}
\usepackage{slashed}
\usepackage{color}
\usepackage{mathrsfs}
\usepackage{subfigure}
\usepackage[font={footnotesize,sl},bf]{caption}
\usepackage{wrapfig}

\usepackage{mathtools}
\usepackage{jheppub}
\usepackage{cancel}
\usepackage{relsize}

\usepackage{enumitem}
\setlist{noitemsep, topsep=.25em, labelindent=1.5\parindent, leftmargin=*} % makes lists a bit more compact

\newcommand\be{\begin{equation}}
\newcommand\bea{\begin{eqnarray}}
\newcommand\ee{\end{equation}}
\newcommand\eea{\end{eqnarray}}

\newcommand{\bdm}{\begin{displaymath}}
\newcommand{\edm}{\end{displaymath}}
\newcommand{\nn}{\nonumber \\}
\newcommand{\f}[2]{\frac{#1}{#2}}
\newcommand{\bref}[1]{(\ref{#1})}
\newcommand\h{\frac{1}{2}}
\newcommand{\ket}[1]{|#1 \rangle}
\newcommand{\bra}[1]{\langle #1 |}

\newcommand{\defeq}{\vcentcolon=}

\newcommand\re{\mathbb{R}}
\newcommand\ints{\mathbb{Z}}
\newcommand\dg{\dagger}
\newcommand{\com}[2]{[#1,\, #2]}
\newcommand{\drm}{\mathrm{d}}

\title{No Holography for Eternal AdS Black Holes}
\author[1]{Steven G.\ Avery}
\author[2]{and Borun D. Chowdhury}
\affiliation[1]
{Department of Physics\\
Brown University\\
Providence, RI 02912, USA
}
\affiliation[2]{Department of Physics \\
Arizona State University \\
Tempe, Arizona 85287, USA
}

\abstract{It is generally believed that the eternal AdS black hole is dual to two conformal field theories with compact spatial sections that are together in a thermofield double state. We argue that this proposal is incorrect, and by extension so are the ``entanglement=geometry'' proposal of Van Raamsdonk and ``ER=EPR'' proposal of Maldacena and Susskind. We show that in the bulk there is an interaction needed between the two halves of the Hilbert space for connectivity across the horizon; however, there is no such interaction between the CFTs. This rules out the possibility of the dual to the CFTs being the eternal AdS black hole. We argue the correct dual ``geometries'' resemble the exterior of the black hole outside the stretched horizon but cap off before the global horizon. This disallows the possibility of a shared future (and past) wedge where Alice falling from one side can meet Bob falling from the other. We expect that in the UV complete theory the aforementioned caps will be fuzzballs.}


\preprint{Brown-HET-1652}

\begin{document}
\maketitle

\section{Introduction}

A standard piece of the AdS/CFT dictionary, first advocated in~\cite{Maldacena:2001kr}, is the
duality between eternal AdS black holes and the thermofield double formalism for finite
temperature conformal field theories. 

We argue here that the proposed duality does not exist except in its most limited form. While the
two asymptotic CFTs do describe the physics outside the black hole's stretched horizons, they do not
describe any black hole interior degrees of freedom. This is not the first argument against a
description of the shared forward wedge of eternal AdS black holes in terms of a thermofield double state of two CFTs (for
instance, see~\cite{Marolf:2012xe,Marolf:2013dba,Avery:2013exa,Almheiri:2013hfa}); however, the arguments
presented here are new and stronger. We argue that there is no extension or completion of the
two CFTs that describes a shared forward wedge that would allow observers from the two asymptotic
regions to meet. 

This has implication for two other proposals as well. 
When studying free or weakly interacting field theories, one finds that the ground state is highly
entangled in position space; in fact, it is ``maximally entangled'', by which we mean as entangled
as is possible given conservation of energy.\footnote{The statement is also expected to hold for
  reasonable, strongly interacting theories. The same is not true in momentum space: the vacuum of
  free field theory in momentum space is entirely
  unentangled. See~\cite{Balasubramanian:2011wt,Hsu:2012gk} for investigations of entanglement in
  momentum space for interacting theories.}  
  %Furthermore, it is clear that all reasonable states, states with finite energy, for instance, share this property. 
  Based on~\cite{Maldacena:2001kr} one might propose the converse:
that entanglement implies a connected space. One might take as a slogan for this point of view
``entanglement=spacetime''. Certainly, this theme runs
through~\cite{VanRaamsdonk:2009ar,Czech:2012bh,VanRaamsdonk:2010pw,VanRaamsdonk:2013sza}.

Further, in response to recent information theoretic arguments against smooth black hole
horizons~\cite{Mathur:2009hf, Almheiri:2012rt, Avery:2012tf,Almheiri:2013hfa},\footnote{An oversight
  in the pivotal argument of~\cite{Mathur:2009hf} was addressed in~\cite{Avery:2011nb}.} Maldacena
and Susskind~\cite{Maldacena:2013xja} proposed a more concrete incarnation of these ideas, which has
been called ``ER=EPR''. (For recent refinements, see~\cite{Susskind:2013lpa, Susskind:2013aaa}.)
The idea is that quantum entanglement (for instance, of the EPR type), can be geometrized as an
Einstein--Rosen bridge. Since these proposals are based on the understanding of how the eternal AdS
black hole is realized in the AdS/CFT correspondence~\cite{Israel:1976ur,Maldacena:2001kr}, we
expect (at the very least) that the simplest interpretations of entanglement=spacetime and ER=EPR
fail to hold.

It is important to note, as emphasized in~\cite{Maldacena:2013xja}, that entanglement is \emph{not}
a good observable; there is no linear, Hermitian operator that measures the entanglement of a
state. The best one can do is check whether the system is in a particular entangled
state.\footnote{For a two-qubit system, this process is called a Bell measurement. We discuss it and
  its relevance to the current work in Appendix~\ref{sec:bell}.} This already suggests that we
proceed cautiously when trying to relate entanglement to spacetime connectivity.

Without further ado, let us consider the eternal AdS black hole. The full spacetime has two
asymptotically AdS regions. The AdS/CFT dictionary suggests that each has its own dual CFT. To put a
single CFT into a thermal state one traditionally puts a black hole in the geometry, and so it is
natural to suppose that one should think of the two-sided black hole as dual to two copies of the CFT
maximally entangled in the so-called thermofield double
state~\cite{Israel:1976ur,Maldacena:2001kr}.

At this point, we would like to emphasize two elementary points:
\begin{enumerate}
\item All Hilbert spaces of the same dimension are isomorphic, so a particular ``state'' is only
  meaningful if we agree what operators we are going to act with before hand.
  \item If Alice is going to measure two degrees of freedom, she must certainly interact with them
  both. In the context of black holes, we can say that moving across the horizon corresponds to interacting with the inside-horizon and the outside-horizon modes and performing a Bell measurement. See Appendix~\ref{sec:bell} for details.
\end{enumerate}
What is the relevance of these points?  Recall the thermofield double formalism, which as its name
suggests involves doubling the degrees of freedom. If we have a thermal state, $\rho \sim
\exp(-\beta H)$, we can imagine purifying the system by adding additional degrees of freedom; the
simplest way is to double every degree of freedom and entangle each degree of freedom in the
thermal system with its partner:
\begin{equation}
\ket{\Psi} = N_\beta \sum_E e^{-\frac{\beta}{2} E}\ket{E}_L\otimes\ket{E}_R.
\end{equation}
One can think of the second copy as modeling the role of the environment.  Our point: this state
does not have a particular meaning unless we agree what operators we act on it with. In particular
the same state can be evolved with different Hamiltonians. For instance, consider the Hamiltonian
\begin{equation}
H_\text{doubled} = H_L + H_R.
\end{equation}
This Hamiltonian has no interaction term between the two copies. This is equivalent to the
standard approximation of ignoring interactions that bring a system to thermality once it is in
equilibrium.
% Since the original CFT in a thermal state only provides observables for one CFT and
%since there is no coupling between the two CFTs, one could perform a general unitary transformation
%on the ``copy'' and the physics of the thermal system should be unchanged; after all, the second
%copy is just mocking up the environmental/heat bath degrees of freedom the original CFT is entangled
%with. This point has been emphasized in this context in~\cite{Marolf:2012xe,
 %Avery:2013exa,Almheiri:2013hfa}.

However, one can also consider Hamiltonians of the form
\begin{equation}
H_\text{doubled-interacting} = H_L + H_R + H_\text{int}.
\end{equation}
which involves an interaction term between the two copies.  We argue that for any quantum or
collection of quanta\footnote{which we might name Alice, for instance.} to pass through the horizon
of the eternal AdS black hole and thereby ascertain whether or not there is a firewall necessitates
an interaction term between the left and right degrees of freedom.  While such an interaction term
is present in the eternal AdS black hole, it is missing between the two CFTs. This shows that the
proposed duality of~\cite{Maldacena:2001kr} does not work.

Below, we outline the major points we make in the paper.

%  We do this by comparing the situation of the eternal AdS black hole to,
%  in turn,
%  \begin{itemize}
%  \item free field theory in 1+1 dimensions arbitrarily broken into left and right halves
%  \item the Rindler description of the same system
%  \item the Rindler AdS/CFT correspondence
%  \end{itemize}
%  This all suggests that the CFT(s) only describes the spacetime outside of the horizon(s). Moreoever,
%  we provide additional evidence that there may not be a sensible way of describing the forward wedge
%  within some extended framework. A slogan for the point of view presented in this paper might be
%  ``entanglement is not enough''.

\subsection{A Toy Model} 

In Section~\ref{FieldTheoryTwain}, we start with a toy problem that demonstrates several of the
above points. We study a free $1+1$ dimensional field theory by arbitrarily breaking it into left
and right halves. 

The first observation we make is that there is an explicit interaction between the two halves,
which ultimately derives from the $(\partial_x\phi)^2$ term in the Lagrangian. This is what allows
excitations to travel from one side to the other.

Our second observation is that one can evolve \emph{the same state with an alternate Hamiltonian}
with a mirror at $x=0$. For instance, one may take the highly entangled global (``interacting'')
vacuum at $t=0$, and evolve it forward with the mirror Hamiltonian. This results in a divergent
pulse of radiation traveling outward along null rays. This ``free'' Hamiltonian does not have the
interaction terms coupling the two halves. Thus, in this field theory example, one can explicitly
see that it is the interactions that connect the two halves, and not entanglement.

Our third observation is that a Rindler observer accelerating away from the origin is completely
ignorant of the potential existence of a fatal light-like pulse. The Rindler coordinates freeze the
interactions at the origin. Moreover, since all correlators are identical in the Rindler wedge for
the mirror and for the non-mirror case, if one analytically continues correlators in the Rindler
wedge to the global Minkowski spacetime, there is no signature of the radiation that will prove
fatal for an inertial observer. Analytic continuation gives an incorrect result.

In the context of holography, the evolution without the mirror corresponds to the bulk AdS black
hole spacetime and we see it involves an interaction between the left and the right. The outside of
the black hole evolves with Schwarzschild time and is the analogue of Rindler wedges. These are
oblivious to this interaction. The CFTs living on the boundary have to be non-interacting on account
of being causally disjoint. If there is some way to extend or complete the CFTs so that they interact
interact, then the dual bulk microstates will also interact, and the AdS black
hole being dual to the CFTs would be consistent. Conversely, if there is no such extension of the CFTs,
then the bulk AdS black hole cannot be dual to the two CFTs. The dual of the CFTs would in the latter
case be analogous to the mirror system, with no connectivity between the left and the right and thus
no shared future or past wedges. We give examples of both kinds in following sections.
    
\subsection{The Rindler-AdS/CFT Correspondence}

In Section~\ref{sec:rindler}, we use the recent Rindler-AdS/hyperbolic CFT correspondence. This
serves as an example where two CFTs can be completed into a global CFT describing a connected
spacetime (global AdS). In this section and the remainder of the paper, we specialize to $AdS_3$,
but we expect our arguments to hold unmodified for higher dimensions.

First, we argue that the two hyperbolic CFTs describe the physics of the two Rindler wedges, but do
not capture the physics behind the Rindler horizons.

Second, we note that the Rindler horizons correspond to the boundaries of the hyperbolic CFTs, so
that connectivity of the bulk gets related to connectivity of the boundary. The two
hyperbolic CFTs are analogous to the Rindler wedges in the toy model; they are ignorant of whether
there are sharp insertions at the connecting point that spoil smooth global AdS. Moreover, the
Hilbert spaces associated with the two CFTs interact if we map back to the global time, just as the
two parts of the global AdS interact when switching away from the Rindler coordinates.

\subsection{The Eternal AdS Black Hole}

In Section~\ref{sec:eternal}, we finally turn to the case of interest, the eternal AdS black
hole. In three bulk dimensions, we have the BTZ black hole, which was also the case explicitly
studied in~\cite{Maldacena:2001kr}.

We recall that the BTZ black hole can be understood as an orbifolding of $AdS_3$. The orbifolding is
particularly simple in Rindler-AdS coordinates. The orbifolding of the bulk translates into an
orbifolding of the hyperbolic CFT, which places the CFT out of contact with the bulk horizon. The
CFT becomes a cylinder instead of an open interval.  

Before orbifolding, an interaction between the two systems can be made manifest by returning to
global AdS in the bulk and the original CFT on a cylinder in the boundary. In the CFT, the
interaction arises just from piecing the two open regions into a complete whole. After orbifolding,
in the bulk one can switch to Kruskal-like coordinates to pass through the horizon so the
interaction is still present; however, since the two CFTs are already on cylinders there is no way
to patch them together and the interaction is turned off.

\subsection{Conclusions}

From the above arguments, we conclude that while two hyperbolic CFTs in a thermofield double state
do have shared future and past wedges in the bulk, two cylindrical CFTs in a thermofield double
state do not, contrary to the proposal in~\cite{Maldacena:2001kr}. Being entangled is necessary, but
not sufficient for connectedness of spacetime. Moreover, the eternal AdS black hole has a smoothly
connected forward wedge, and relatedly has quasi-normal modes. The CFT on a compact space has no
quasi-normal modes.\footnote{Technically quasi-normal modes are defined for the bulk. In the CFT the
  signature of bulk quasi-normal modes are poles of the retarded 2-point function in the lower half
  complex plane in momentum space. In this paper we refer to a CFT having such complex poles as
  having quasi-normal modes.} This suggests that the two CFTs in a thermofield double state is dual
to the left and right wedges of the eternal black hole cut off at the stretched horizons. There is
no interior description, although we expect the UV complete description of the geometry to replace
the stretched horizon by fuzzball microstates.

\section{Breaking a Field Theory in Twain} \label{FieldTheoryTwain}

As a demonstration of the key ideas, we introduce the following toy problem. We consider breaking
$1+1$-dimensional field theory into (strongly interacting) left and right halves. We explicitly
demonstrate, in case there was any doubt, that the connectivity of the space derives from an
interaction term. Moreover, we consider evolving the original vacuum state with the noninteracting
Hamiltonian, which results in a divergent stress tensor propagating outward from the break
point. This is analogous to a firewall along the Rindler horizon of an accelerating observer.

In the context of holography, this example illustrates that for observers in different asymptotic
regions of the eternal AdS black hole to meet in the future, there must either be an interaction
term directly between the two CFTs or between each CFT and a common third system. Neither of which
is the case in the standard thermofield double formalism.  At which point, one may consider trying
to extend the CFT side of the duality in some way to describe a shared future, or one may consider
altering the bulk side of the duality so that the two asymptotic AdS regions are not connected. 

\subsection{Global aspects}

Consider a free, massless scalar field with Dirichlet boundary conditions:
\begin{equation}
\mathcal{L} = \frac{1}{2}(\dot{\phi}^2 - {\phi'}^2)\qquad \phi(t, -\tfrac{L}{2}) \equiv \phi(t,\tfrac{L}{2}) \equiv 0\quad\forall t\in \re.
\end{equation}
We put the field theory in a box as a convenient IR regulator. The field can be expanded in modes
\begin{equation}
\phi_m(t, x) = \frac{1}{\sqrt{\omega_m L}} \sin\left(\omega_m (x+ \tfrac{L}{2})\right)e^{-i\omega_m t}\qquad \omega_m = \frac{m\pi}{L}\quad m\in\ints
\end{equation}
in the usual way
\begin{equation}
\phi(t, x) = \sum_{m=1}^\infty\left(a_m\phi_m + a_m^\dg \phi_m^*\right)\qquad \com{a_m}{a_n^\dg} = \delta_{m,n}.
\end{equation}

We are interested in now formally breaking the field into a left and right field ($x$ negative and
positive, respectively):
\begin{equation}
\phi(t, x) = \begin{cases}
\psi(t, x) \qquad & -\tfrac{L}{2}\leq x < 0\\
\chi(t, x) & 0< x \leq \tfrac{L}{2}
\end{cases}
\end{equation}
Let $\psi(t,x)$ be the left field with modes $b_m$ and $\chi(t,x)$ be the
right field with modes $c_m$. We arbitrarily impose Dirichlet boundary conditions on both fields at $x=0$.
The expansions are given by
\begin{equation}\begin{aligned}\label{eq:psi-chi}
\psi_m &= \sqrt{\frac{2}{\kappa_m L}} \sin(\kappa_m x)e^{-i\kappa_m t}\qquad \kappa_m = \frac{2m\pi}{L}\\
\chi_m &= \sqrt{\frac{2}{\kappa_m L}} \sin(\kappa_m x)e^{-i\kappa_m t}.
\end{aligned}\end{equation}
These expansions can be thought of as in the interaction picture, where the time-dependence due to
the interaction between the two halves has not been included.

We are left with two field expansions that we equate at $t=0$:
\begin{equation}
\sum_{m=1}^\infty\left( a_m \phi_m + a_m^\dg \phi_m^*\right)
 = \begin{cases}
\sum_{n=1}^\infty \left( b_n\psi_n + b_n^\dg \psi_n^*\right) \qquad & x <0\\
\sum_{n=1}^\infty \left( c_n\chi_n + c_n^\dg \chi_n^*\right) & x > 0
\end{cases}.
\end{equation}
Using the orthogonality of the mode functions we may then extract the Bogolyubov coefficients
relating the different modes:
\begin{equation}\begin{aligned}
a_{2k} &= \frac{(-1)^k}{\sqrt{2}} (b_k + c_k)\\
a_m &= \frac{(-1)^\frac{m-1}{2}}{L} \sum_{n=1}^\infty\sqrt{\frac{2\kappa_n}{\omega_m}}\left[
\frac{b_n - c_n}{\omega_m - \kappa_n}
+\frac{b_n^\dg - c_n^\dg}{\omega_m + \kappa_n}\right]
\qquad (m\,\text{odd}).
\end{aligned}\end{equation}
The mixing between the lowering operator and raising operators implies that the notions of vacua do
not agree. It turns out, one can write
\begin{equation}\label{eq:a-vacuum}
\ket{0_a} = \prod_{m,n} \exp\left[\frac{1}{2}\gamma_{m,n}(b_m^\dg-c_m^\dg)(b_n^\dg-c_n^\dg)\right] \ket{0_b}\ket{0_c},
\end{equation}
with
\begin{equation}
\gamma_{m,n} = \frac{2\sqrt{\kappa_m\kappa_n}}{L^2}
   \left(\sum_{k\in\text{odd}^+}\frac{1}{\omega_k(\kappa_m - \omega_k)(\kappa_n + \omega_k)}\right)
 = \frac{2\sqrt{\kappa_m\kappa_n}}{L^2} s_{m,n}.
\end{equation}
This equation implies, as expected, that the $a$-vacuum is highly entangled between the left and
right halves.

One may be concerned with this decomposition, since in the $\psi$/$\chi$ description we have imposed
Dirichlet boundary conditions at $x=0$ where our original theory had no such condition. We expect
that generically this is not a problem since the fields $\psi$ and $\chi$ can converge to any $\phi$
in an $L^2$-norm sense, for instance. Some issues may arise if operators are inserted exactly at
$x=0$, as became evident in a related set-up explored by the current authors~\cite{Avery:2010vk}.

Let us proceed to rewrite the original Hamiltonian,
\begin{equation}
H_a = \sum_m \omega_m \left(a_m^\dg a_m + \tfrac{1}{2}\right) \label{BigBoxHamiltonian}
\end{equation} 
in terms of the left and right modes. The number operator is straightforward to write
out. For the even modes
\begin{equation}
a_{2m}^\dg a_{2m} = \frac{1}{2}\left(
b_m^\dg b_m + c_m^\dg c_m + b_m^\dg c_m + c_m^\dg b_m\right).
\end{equation}
Note that the latter two terms already have the form of an interaction which ``moves'' degrees of
freedom between the left and right sides. The first two terms are just the ``free'' Hamiltonian
terms for the left and right seperately.  For odd modes
\begin{multline}
a_m^\dg a_m = \frac{2}{\omega_m L^2}\sum_{k,l=1}^\infty\sqrt{\kappa_k\kappa_l}\bigg[
\left(\frac{1}{(\omega_m-\kappa_k)(\omega_m-\kappa_l)}+\frac{1}{(\omega_m+\kappa_k)(\omega_m+\kappa_l)}\right)
(b_k^\dg - c_k^\dg)(b_l-c_l)\\
+\frac{(b_k^\dg-c_k^\dg)(b_l^\dg-c_l^\dg)}{(\omega_m-\kappa_k)(\omega_m+\kappa_l)}
+\frac{(b_k-c_k)(b_l-c_l) }{(\omega_m+\kappa_k)(\omega_m-\kappa_l)}\bigg]
+ \epsilon_m,
\end{multline}
where
\begin{equation}
\epsilon_m = \frac{4}{L^2\omega_m}\sum_{k=1}^\infty \frac{\kappa_k}{(\omega_m+\kappa_k)^2}
\end{equation}
is the (divergent) $\phi_m$ mode contribution to the difference between the $bc$ vacuum energy and
the $a$ vacuum energy.

Our purpose here is to explicitly exhibit the interaction term between the left and right halves of
the theory. While we can think of the $b^\dg c^\dg$ and $bc$ type terms as being related to the
shift in the vacuum, the $b^\dg c$ and $c^\dg b$ type terms are related to the ability for
excitations on the left to travel to the right and vice-versa. It is precisely this kind of
interaction term that equilibrates between the left and right halves, and allows for
the two otherwise disjoint systems to be entangled in the first place.

In particular, we can compare the Hamiltonian $H_a$ discussed above, to the natural Hamiltonian for
the left and right halves with Dirichlet boundary conditions,
\begin{equation}
H_\text{mirror} = H_b + H_c = \sum_{k=1}^\infty \kappa_k (b_k^\dg b_k + c_k^\dg c_k), \label{SmallBoxHamiltonian}
\end{equation}
which has no interaction terms. This theory corresponds to inserting a two-sided mirror at $x=0$ of
the original system. It is clear that excitations in this system cannot travel between the two
halves. This is the ``free'' Hamiltonian which gives the time dependence of the interaction picture
modes $\psi$ and $\chi$ in Equation~\eqref{eq:psi-chi}.

While we have a distinct Hamiltonian with a new vacuum $\ket{0_b}\ket{0_c}$, there is no obstacle to
considering the same highly entangled state in Equation~\eqref{eq:a-vacuum} or small excitations
on top of it. Suppose, for example, we suddenly insert a mirror at $x=0$ and $t=0$. At $t=0$, we
have the same state.\footnote{By ``suddenly'', we mean fast compared to the time-scale set by the UV
  cutoff, so that there is no time for the system to react.}

This is why we emphasize in the Introduction that we need to be clear what operators we are using on
our states. The state~\eqref{eq:a-vacuum} of itself cannot tell us whether there is a mirror or
not. This is determined by which Hamiltonian we use: $H_a$ or $H_\text{mirror}$. That is to say, we
could move from a Schrodinger picture at $t=0$, and then evolve with whichever is the physically
relevant Hamiltonian.

\begin{figure}
\begin{center}
\subfigure[Plot of $T_{tt}$ at time $t$]{\label{fig:Ttt}\includegraphics[width=8.5cm]{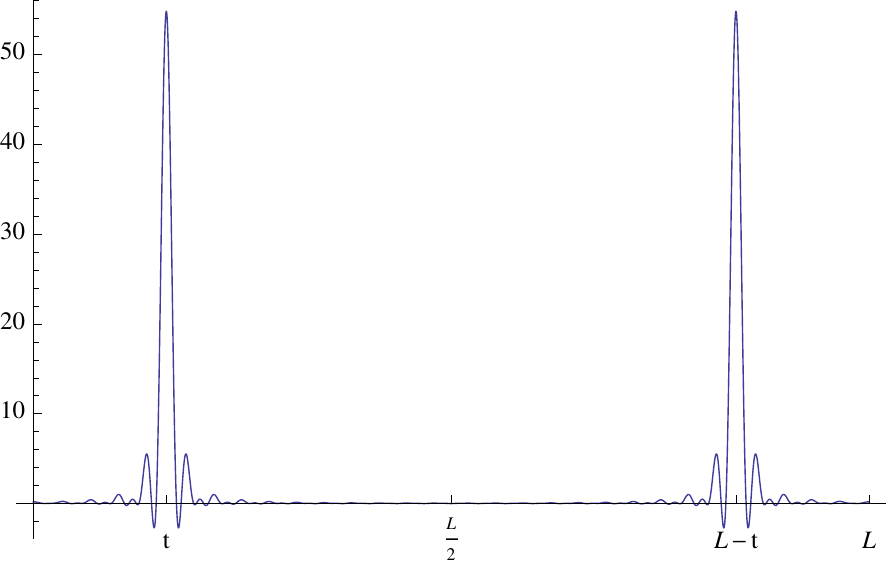}}\hfill
\subfigure[An accelerated observer]{\label{fig:acc-obs}\includegraphics[width=5cm]{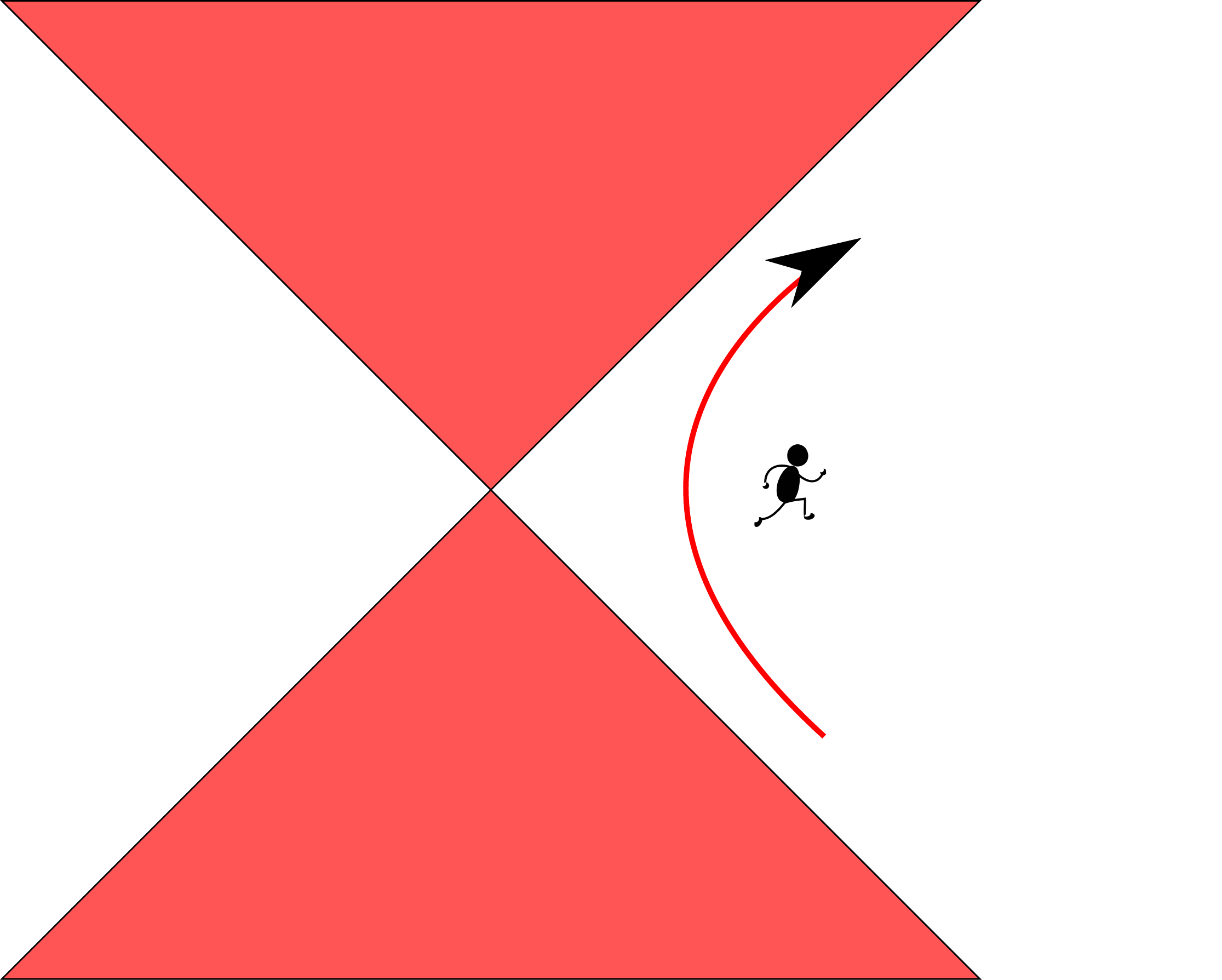}
}
\end{center}
\caption{In~\subref{fig:Ttt}, $T_{tt}$ is plotted at time $t<L/2$ after the insertion of the
  mirror. We cut off the sums at $m,n=30$ in Equation~\eqref{eq:Ttt}. The left pulse is created at
  the mirror ($x=0$) and propagates away along the lightcone.  Note that box wall is at
  $x=L/2$. The pulse at $x=L-t$ is actually the left-moving bounced pulse that only enters the
  physical region once the pulse leaving the quench site at $x=0$ hits the
  wall. In~\subref{fig:acc-obs}, an accelerated observer escapes the divergent stress-tensor, and
  does not notice that a mirror was inserted.}
\end{figure}

What happens to the example under discussion for $t>0$? Inserting the mirror can be thought of as a
local quench similar to the kind studied in~\cite{Asplund:2011cq}. Thus, we expect excitations
to travel outward causally from the quench site. In the current case of a massless field, this means
excitations traveling outward along the light cone. Let us compute the energy density of this state
using the mirror Hamiltonian. The energy density is given by
\begin{equation}
T_{tt} = \mathcal{H}  =\frac{1}{2}(\dot{\phi}^2 + {\phi'}^2),
\end{equation}
and we want to evaluate the difference from the $bc$ vacuum:
\begin{equation}
t_{tt} = \bra{0_a}T_{tt}\ket{0_a} - \bra{0_{b,c}}T_{tt}\ket{0_{b,c}}.
\end{equation}
Let us just focus on $x>0$. After some manipulation, 
\begin{multline}\label{eq:Ttt}
t_{tt} = \frac{4}{\pi^3}\sum_{m,n=1}^\infty\kappa_m\kappa_n\biggr[\tilde{s}_{m,n}\cos\big((\kappa_m-\kappa_n)x\big)\cos\big((\kappa_m-\kappa_n)t\big)\\
+s_{m,n}\cos\big((\kappa_m+\kappa_n)x\big)\cos\big((\kappa_m+\kappa_n)t\big)\biggr].
\end{multline}
Both $s_{m,n}$ and $\tilde{s}_{m,n}$ can be expressed in terms of harmonic numbers:\footnote{We
  remind the reader that the harmonic numbers for integer $n$ can be expressed as $H_n =
  \sum_{k=1}^n \frac{1}{k}$ or for general $n$ as $H_n = \int_0^1\drm x\frac{1-x^n}{1-x}$.}
\begin{subequations}
\begin{align}
s_{m,n} &= \frac{m(2H_{2n}-H_n)+n(2H_{2m}-H_m)}{8mn(m+n)}\\
\tilde{s}_{m,n} &= \frac{m(2H_{2n}-H_n) - n(2H_{2m}-H_m)}{8mn(m-n)}.
\end{align}
\end{subequations}
For $m=n$, one must take the limit as $m\to n$ in the above expression for $\tilde{s}$; the limit is
finite and well-defined. The above expressions are manifestly symmetric in $m$ and $n$, and for
$m\neq n$ are rational, positive numbers. The expression for the stress-tensor~\eqref{eq:Ttt}
diverges. We can regulate the series by putting in a cutoff, $m,n\leq\Lambda$. The stress-tensor is
plotted for $\Lambda = 30$ in Figure~\ref{fig:Ttt}. As $\Lambda\to\infty$, $t_{tt}$ tends to zero
away from the pulse emitted by the insertion site, whereas the peak diverges like
$\Lambda^2\log\Lambda$. The scale of the $\log$ is presumably given by the insertion timescale.

\subsection{Rindlerization freezes interactions}

We saw in the previous section that we can evolve the vacuum state of the box with Dirichlet
boundary conditions at $x=\pm L/2$ (i.e. the vacuum of the hamiltonian~\eqref{BigBoxHamiltonian})
with the hamiltonian corresponding to having a mirror at $x=0$ (i.e. the
hamiltonian~\eqref{SmallBoxHamiltonian}) starting at $t=0$. This results in a divergent stress
tensor, which moves outwards from $x=0$ and $t=0$ towards the two walls of the original box.  The
pulse of radiation eventually bounces off the walls; however, we introduced the box only as a
convenient IR regulator. Therefore we are interested in the physics for $x, t\ll L$.

Taking the limit of $L \to \infty$, the original state becomes the Minkowski
vacuum. We do not have to
worry about how the quench is inserted at $t=0$. Instead, we can imagine the initial state at $t=-\infty$
was created such that the evolution with the mirror hamiltonian produced the Minkowski vacuum at
$t=0$. Or we can simply be indifferent to how the state was created at $t=0$ and evolve it with the
mirror hamiltonian. If we introduce massive fields then  the divergent
stress tensor would also be present in the future and past wedges. The resulting scenario is shown in
Figure~\ref{fig:acc-obs}.

As is clear from the figure, an accelerated observer forever stays away from the divergent stress
tensor. In fact, the \emph{local state} seen by the accelerated observer is the Minkowski vacuum
although the global state will be changing with time. More formally, if we Rindlerize the state in
the usual way, the Rindler wedges for the mirror hamiltonian will be identical to Rindler wedges for
the Minkwoski hamiltonian. In particular correlation functions with all points inside the Rindler
wedge will be identical. Correlation functions with points outside will be different in
general.\footnote{It should be noted that we are talking about Rindlerization with a mirror on the
  state identical to the Minkwoski vacuum at $t=0$. The case with the vacuum state of the
  mirror hamiltonian is not relevant for us but was recently studied in~\cite{Rovelli:2011ur}.}

The Rindler coordinates natural to the accelerating observer slice the spacetime in a peculiar
way. As one approaches the coordinate singularity along the horizon, time steps forward by smaller
and smaller increments until it stops and the coordinates degenerate. Thus any left--right
interactions become frozen. In order for the mirror to leave no imprint on both Rindler wedges, one
requires that the mirror be infinitesimally thin and inserted instantaneously, as compared to the
theory's UV cutoff. In what follows, we argue the accelerating observer is analogous to a single
CFT's description of the eternal AdS black hole, with some important differences related to the
mirror ``thickness''. 

We hope the relevance of this toy problem to the firewall discussion is clear: the accelerated
observer cannot tell whether the system evolves with $H_a$ from \bref{BigBoxHamiltonian}, allowing
``free infall'' via its required interaction terms, or with $H_\text{mirror}$ from
\bref{SmallBoxHamiltonian}, corresponding to a ``firewall''. Let us emphasize that the entanglement
between left and right halves is the \emph{same} in both situations. Within field theory, when there
is no dynamical theory of gravity, entanglement does not imply a smoothly connected spacetime. Using
AdS/CFT, we extend the argument into the bulk.

We can also use this example to address the issue of analytic continuation. One can imagine
analytically continuing a Green's function in the right Rindler wedge onto the whole Minkowski
space; however, since all correlators are identical on the wedge with and without the mirror,
the existence of a smooth analytic continuation does \emph{not} imply a smooth Rindler horizon.

\section{Rindler-AdS}\label{sec:rindler}

\subsection{Review of Rindler-AdS}

\begin{figure}[htbp]
\begin{center}
\subfigure[Bulk AdS with acceleration horizons and Bulk-Alice falling in. $\tau$ increases in the vertical direction and $\phi$ increases anti-clockwise.]{
\includegraphics[scale=.5]{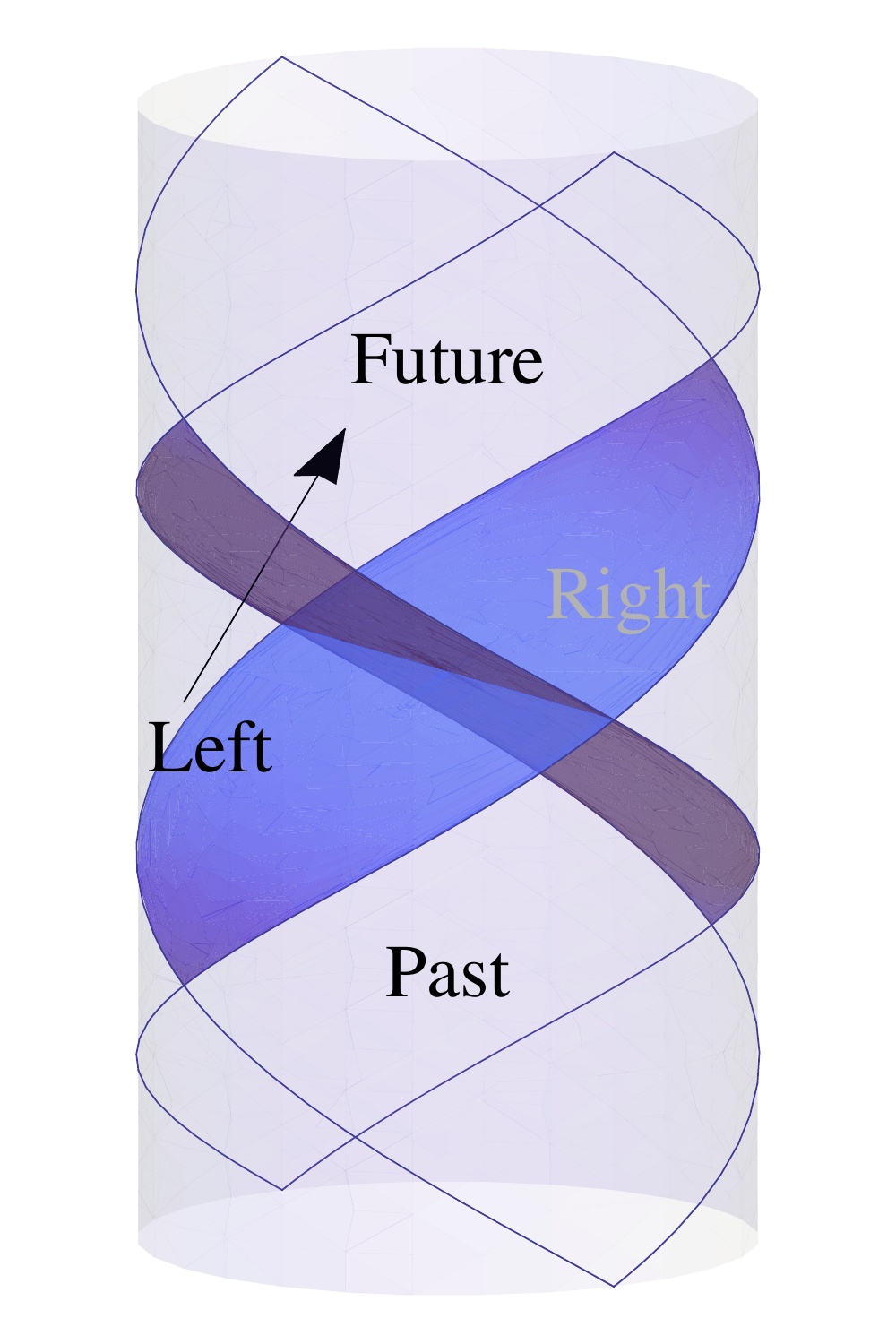} \label{RindlerAdSBulk}} \hspace{1cm} 
\subfigure[Causal diamonds on global boundary with Boundary-Alice escaping. $\tau$ increases in the vertical direction and $\phi$ increases towards the right.]{
\includegraphics[scale=.5]{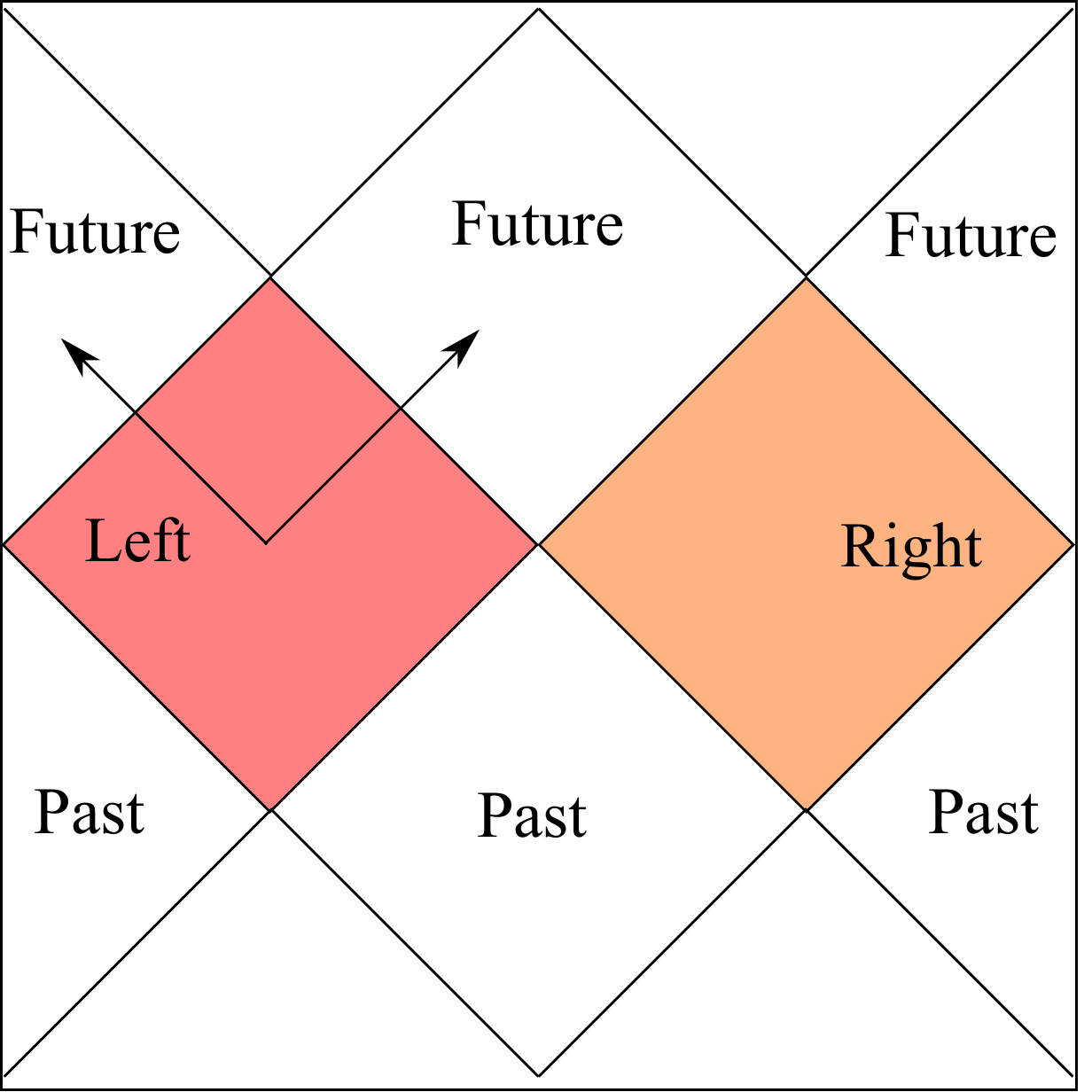} \label{RindlerAdSBoundary}}
\caption{Global AdS can be ``Rindlerized''. In (a) the associated acceleration horizons and closed string Bulk-Alice falling through one of them is shown. In (b) the associated causal diamonds and open string Boundary-Alice escaping it is shown.}
\label{RindlerAdS}
\end{center}
\end{figure}

Global $AdS_{d+1}$ space can be ``Rindlerized'' pretty much like flat spacetime resulting in bulk
acceleration horizons~\cite{Emparan:1999gf,Czech:2012be,Parikh:2012kg}. To simplify our
discussion, we focus solely on the case of $AdS_3$. The Carter--Penrose diagram of $AdS_3$ with
acceleration horizons is shown in Figure~\ref{RindlerAdSBulk}. These ``black holes'' have horizons,
but no singularities as is evident from the figure.

The metric of global $AdS_3$ is is given by 
\be 
ds^2 = \sec^2 u (-d\tau^2 + du^2 + \sin^2 u ~d \phi^2) \label{GlobalAdS} 
\ee 
where the coordinate $u \in [0,\pi/2)$. These coordinates are helpful in visualising global AdS as
conformal to a finite radius cylinder. According to AdS/CFT duality, the dual CFT is defined on a
background geometry given by the conformal boundary of the AdS space; in this case $\re \times S^1$.

The global coordinates, $\tau$, $u$ and $\phi$, can be exchanged for Rindler-AdS coordinates,
$t$,$r$ and $\chi$. The details are given in Appendix~\ref{variousCood}. The metric then becomes 
\be
ds^2 =- (r^2-1) dt^2 + \f{d r^2}{r^2-1} +r^2 d \chi^2 \label{TopologicalBlackHole} 
\ee 
where $\chi \in (-\infty,\infty)$. Note that the acceleration horizon is at $r=1$ and the
coordinates cover only one of the Rindler-AdS wedges shown in Figure~\ref{RindlerAdSBulk}. There is
a second set of coordinates that covers the other Rindler-AdS wedge.  As is obvious from the large
$r$ behaviour, the dual CFT is defined on $\re^{1,1}$.

The acceleration horizons intersect the boundary cylinder at the edges of causal diamonds. The
inside of these diamonds describe the causal development of the open intervals $\phi \in (-\pi/2,
\pi/2)$ and $\phi \in (\pi/2, 3 \pi/2)$ at $t=0$ (see Figure~\ref{RindlerAdSBoundary}).

It was shown in~\cite{Casini:2011kv} that the conformal transformation,
\be
\tan (\tau) = \f{\sinh (t)}{\cosh (\chi)}, \qquad \tan (\phi) = \f{\sinh (\chi)}{\cosh (t)}, \label{CFTmap}
\ee
maps the causal development of the interval $\phi \in (-\pi/2,\pi/2)$ with metric
\be
ds^2 = -dt^2 + d\phi^2 
\ee
into  a space conformally related to $\re^{1,1}$:
\be\label{eq:weyl-mink}
ds^2 =  \f{2}{ \cosh (2 \chi) + \cosh (2 t) } (-dt^2 + d \chi^2).
\ee
Under this map, the vacuum of the full CFT gets mapped into a thermal state on the
space~\eqref{eq:weyl-mink}. The coordinates $t,\chi \in\re$, thus making the causal diamond a
complete theory by itself.  This is a purely field theory result, but the map can be seen as a
projection of the bulk coordinate transformation \bref{CoordRelb} and \bref{CoordRelc}. The space
and time dependent conformal factor comes from \bref{CoordRela} and captures the effect of mapping
the CFT on constant $r$ surfaces to that on constant $u$ surfaces. In higher dimensions the CFT is
defined on hyperbolic spacetime~\cite{Emparan:1999gf,Czech:2012be,Parikh:2012kg}, so we
will refer to the two dimensional CFT on~\eqref{eq:weyl-mink} as the hyperbolic CFT.

\subsection{How much does a hyperbolic CFT capture?}

So far, we have only discussed the global vacuum state. What we are really interested in is
tracking a bulk excitation passing through the acceleration horizon and interpreting this motion in
the CFT. To this end, we ask what is the region of bulk spacetime which is uniquely fixed by the
state on a Cauchy slice which covers only a part of the boundary cylinder? For small enough
perturbations initially localised in the causal diamond of the Cauchy slice we can then say that the
dual bulk excitations are completely determined by the associated hyperbolic CFT as long as they are in
the corresponding bulk region.  Recall that since the CFT is local, one can write the complete
Hilbert space in a product form
\be
H = H_A \otimes H_{\bar A},
\ee
where $H_A$ is associated with the relevant Cauchy slice and $H_{\bar A}$ with its complement. So
the question is given a state on $H_A$ how much of the bulk is fixed.

If the map between boundary degrees of freedom and bulk spacetime is sufficiently non-local it could
happen that data from the entire boundary is needed to construct any subregion of the bulk. However,
there are reasons to believe that the situation is not so bad.

One reason to expect a subregion duality is the following. If we do not take the decoupling limit,
then the core region of a stack of D-branes is AdS-like and a closed string tunnelling through the
graybody factors and falling in has an alternate description as the closed string hitting the stack
of branes and becoming open
strings~\cite{Callan:1996dv,Dhar:1996vu,Das:1996jy,Das:1996wn,Maldacena:1996ix,Lunin:2001dt,Chowdhury:2007jx,Avery:2009tu}. The
essential idea of AdS/CFT is that in the decoupling (low-energy) limit, the core geometry becomes
asymptotically AdS and the D-brane theory flows to a CFT. When the closed string is close to the
boundary, the open strings are close to each other and the action of ``plucking'' the D-branes is
very localised. As the closed string goes deeper, the open strings spread
out~\cite{Lunin:2001dt}. Thus, it seems likely that when the bulk excitation is
close to the boundary, it is captured by local operators in the boundary.

Another reason to expect a subregion duality is that the asymptotic behavior of the fields in the
bulk is directly related to the expectation values of local operators in the boundary field theory
(together with the field theory action). One expects that using the boundary behavior of the bulk
fields in some region and the bulk field equations, one may integrate
the field equations into some neighbourhood of the boundary
subregion~\cite{Czech:2012bh}.

Intuition from AdS/CFT and causality would suggest that the boundary data should in the very least
be enough to construct the Rindler-AdS wedge corresponding to the boundary causal diamond. Such a
picture has indeed been advocated in~\cite{Czech:2012bh,Hubeny:2012wa,Bousso:2012mh}. We work
under that assumption.

Now imagine introducing an excitation called Alice into the system. We imagine that bulk-Alice,
made of closed strings, is introduced very close to the boundary in the center of some causal
diamond as shown in Figure~\ref{RindlerAdSBulk}. From the above discussion, the boundary-Alice will
be made of left- and right-moving open strings that are entangled in general. Further, from the
above discussion we expect the motion of bulk-Alice while she is in the Rindler wedge to be
captured by the hyperbolic CFT living in the causal diamond as shown in
Figure~\ref{RindlerAdSBoundary}.

When bulk-Alice crosses the acceleration horizon, boundary-Alice can no longer be described solely
by the hyperbolic CFT she was created in since that hyperbolic CFT can only uniquely determine the
bulk up to the acceleration horizon. When bulk-Alice crosses her future horizon, she might be hit by
any excitation created in the other Rindler wedge---say bulk-Bob---in the region marked ``Future'' in
Figure~\ref{RindlerAdSBulk}. Similarly, boundary-Alice after coming out of the causal diamond may
get hit by the dual to bulk-Bob---boundary-Bob who was created in the antipodal hyperbolic CFT---in
the region marked ``Future'' in Figure~\ref{RindlerAdSBoundary}.

More formally, even without the details of the construction, it is clear that a bulk two-point
function with both points inside the Rindler-AdS wedge, corresponding to the propagator that keeps
bulk-Alice inside the Rindler-AdS wedge, can be written in terms of $n$-point functions of the
corresponding hyperbolic CFT. However, the bulk two-point function with one point inside the
Rindler-AdS wedge and the other in the forward wedge, corresponding to the propagator that moves
bulk-Alice across the acceleration horizon, has to be written in terms of $n$-point functions
with points outside the hyperbolic CFT. In this sense, the global CFT is the
theory of the inside of the Rindler-AdS ``black hole''.

Note that the hyperbolic CFTs cannot describe boundary-Alice and boundary-Bob leaking out since
the boundary time tends to infinity at the edges. In fact the hyperbolic CFTs are defined on open
intervals and do not include the boundaries. Recall that the boundaries of the diamond are
projections of the bulk acceleration horizons. The Rindler-AdS bulk spacetimes also cannot describe
bulk-Alice and bulk-Bob crossing the horizons since Rindler time tends to infinity at the
horizons. Switching to global coordinates in the bulk and the boundary we see that Alice and Bob
have no trouble crossing the respective horizons and meeting in the bulk and similarly the boundary
versions come out of the causal diamond and meet.

\subsection{What is dual to the horizon?}

Let us discuss if the two hyperbolic CFTs together are equivalent to the global CFT.  At $t=0$ they
together span the open interval $(-\pi/2,\pi/2) \cup (\pi/2, 3\pi/2)$ and thus together the two CFTs miss the
points $\phi=\pm\pi/2$; however, since these are just points one may wonder how much Cauchy
data is missing. The correct way to handle this is to introduce UV cutoffs in the global CFT so that
the point has some finite thickness. For the map~\bref{CFTmap} it can be shown that the interval
$(\pi/2 - \delta\phi,\pi/2)$ maps to $(\chi_\text{max}, \infty)$ where 
\begin{equation}
e^{-\chi_\text{max}}  \approx \frac{\delta \phi}{2\cosh t}.
\end{equation}
In ref.~\cite{Casini:2011kv} this was used to argue that a UV cutoff in the global CFT is related to
an IR cutoff in the hyperbolic CFT. Thus, if we remove the IR cutoff on the two hyperbolic CFTs they
indeed are together equivalent to the global CFT because the two missing points are of zero
measure. There is no room to insert a ``reasonable'' operator at $\phi=\pm\pi/2$.  However, one
can imagine introducing a very sharp operator at the point where the two causal diamonds meet and
this will propagate along the forward edges of the causal diamonds in the boundary. In the bulk the
corresponding signal will move along a null path and depending on the precise operator may collide
with Alice crossing the acceleration horizon. So, the acceleration horizons in the bulk are only
captured by the hyperbolic CFTs in the absence of any IR cutoff. This will be crucial when we
discuss getting the BTZ black hole by orbifolding $AdS_3$.\footnote{Points in the horizon are
  contained within many different Rindler-AdS wedges so for any given point asking which subregion
  of CFT is dual to it does not make sense. However, once we have chosen specific Rindler-AdS wedges
  we can ask the question of what is dual to the horizon.}

%We discussed that the minimum region of bulk spacetime constructed from a subregion of the boundary should be the associated Rindler-AdS wedge. However, another interesting question for our purposes is to ask what captures the acceleration horizon in the CFT. From the discussion above recall that the edges of the causal diamond are the projection of the bulk acceleration horizons. The hyperbolic CFTs are defined on the inside of the causal diamonds and therefore do not account for the evolution of the edges. However, we also saw how these edges are of zero measure and removing IR cutoffs from the CFTs meant that the two hyperbolic CFTs completely capture the global CFT. So it stands to reason that the two CFTs together capture the acceleration horizon also.

%However, this must be taken with a pinch of salt. If we keep an IR cutoff in the hyperbolic CFTs then the horizons are not captured by them. This is because an IR cutoff in the hyperbolic CFTs corresponds to a UV cutoff in the global CFT and then one can squeeze in a sharp enough operator between the two causal diamonds which makes the edges of the diamonds non-smooth in the boundary and in the dual setup makes the acceleration horizons non-smooth. 

\subsection{Signatures of connectivity in the CFT} \label{sec:Connectivity}
 
We saw that connectivity in the bulk is associated with connectivity in the boundary: when the two
Rindler-AdS wedges are secretly part of global AdS, the hyperbolic CFTs are part of the global CFT. In
the general relativistic description of the bulk, there are clear diagnostic signals indicating that
one can extend the solution past the bulk horizon: the geometric invariants are all smooth and
finite at the horizon, and timelike geodesics reach it in a finite amount of proper time. Are there
corresponding indications of the ``complete-ability'' of the hyperbolic CFT? Here we are interested
in isolating the conditions in the CFT that led to connectivity/complete-ability, or in other words
allowed Alice, a creature created inside one of the causal wedges, to escape unharmed and possibly
meet Bob who has been created in the antipodal causal wedge. Below we list these conditions.

\begin{itemize}
\item\textbf{Interaction:} As discussed earlier, the two hyperbolic CFTs are complete theories by
  themselves and are in particular non-interacting.  It would seem counterintuitive that excitations
  in two non-interacting theories can meet. Let us begin by discussing how this comes about. Recall
  from Section~\ref{FieldTheoryTwain} that Rindlerization makes the theory in the Rindler wedge
  indifferent to the nature of interaction, or lack thereof, between two halves of the system. This
  is  because their inner boundaries---the acceleration horizons---move away at speed of
  light to escape any effect of an interaction between the two halves.  This is also true of
  Rindlerization of AdS. Of course, the two halves of global AdS do interact. Similarly, though the
  two halves of the cylinder $\phi \in (-\pi/2,\pi/2)$ and $\phi \in (\pi/2,3\pi/2)$ do interact for
  all $\tau$, the way the hyperbolic CFTs are defined---only inside the causal diamond---they are
  ignorant of this interaction. Thus, while the two hyperbolic CFTs are non-interacting, it is via a
  carefully tuned evolution of initial Cauchy data with ``ever slowing time'' that this is
  arranged. Connectivity actually comes from the two sub-Hilbert spaces, corresponding to the two
  halves of the cylinder, interacting. What allows this interaction is of course the fact that the
  global CFT lives on a connected manifold. It is precisely this interaction which causes the
  entanglement of the two sub-Hilbert spaces in the first place. So, interaction, entanglement and
  connectedness go hand in hand in this example and we cannot really say one causes the other.
\item {\bf Subregion CFTs are open systems:}
Alice can escape the causal diamond and
possibly meet Bob only if the propagators on the global boundary are vacuum
propagators. Consider the two-point function of an operator of dimension $(1,1)$ on the vacuum
state on the cylinder,
\begin{equation}\label{eq:twopt-cyl}
 \langle O(\tau, \phi) O(0,0) \rangle =\f{1}{16}~ \f{1}{\sin^2(\f{\tau- \phi}{2 })\sin^2(\f{\tau+ \phi}{2 })},
\end{equation}
where we have taken the spatial extent of the CFT to be $2 \pi$. Using the conformal
map~\bref{CFTmap} the two-point function for the hyperbolic CFT becomes
\begin{equation}\label{eq:twopt-hyp}
\langle O(t,\chi) O(0,0) \rangle =\f{1}{16}~ \f{1}{ \sinh^2( \f{ t-\chi}{2}) \sinh^2( \f{ t+\chi}{2})} 
\end{equation}
which is the two-point function on the infinite line with temperature $\f{1}{2\pi}$. Reversing the
logic, given a thermal two-point function for a CFT on an infinite spatial line, we can conformally
compactify the associated causal diamond and embed it in a bigger space in various ways---we can
have more than two causal diamonds or we can have one big and one small one---and analytically
continue the two-point function. The maps in the other cases are of course different from that in
\ref{CFTmap}. Analytic continuation only gives the correct answer if the state near the edges of the causal diamond is the vacuum state with respect to the global time of the particular map in question. From the discussion earlier about the relation between UV cutoffs of the global
CFT and IR cutoffs of the hyperbolic CFT, we see that for connectivity, hyperbolic CFTs have to be
in thermal states to arbitrarily \emph{low} energy scales. Said differently, this means that there
is no IR cutoff and in particular there is no mass gap. This is related to the conformally compactified  intervals in the global CFT being open. A possible answer to the question of ``complete-ability'' of
the CFT could simply be that only when the CFT is hyperbolic and thermal to arbitrarily low energies
can it be completed since that implies it is an open interval in the vacuum state. This answer
seems trivial and contradicts the proposal in~\cite{Maldacena:2001kr}; however, we argue against
parts of the duality for eternal AdS black holes below. This answer may be somewhat disappointing
because it simply says the bulk is completable and therefore connected when the boundary is too.
\item {\bf Quasi-normal modes:} Thermal CFTs on infinite volume have quasi-normal modes. In fact, in
  the case of $1+1$ dimensional CFT at finite temperature and infinite spatial extent, the
  quasi-normal modes are fixed by conformal invariance~\cite{Birmingham:2001pj}. We saw that the
  vacuum on open intervals is mapped to a thermal state on infinite intervals. Thus if the previous
  conjecture about ``complete-ability'' is true, quasi-normal modes in the CFT are a necessary (but
  not sufficient) condition for smooth horizons in the bulk. 
\item {\bf Trans-horizon excitations:}
Another signature of connectivity, related to the preceding discussion, is the possibility of
excitations which do not fit into any of the individual intervals. From the point of view of the
hyperbolic CFTs these are analogous to ``trans-horizon excitations''. In fact, even in the bulk they
correspond to transhorizon excitations.
\item {\bf Inhomogeneity:}
The previous point is also related to an inhomogeneity coming from the fact that when we look at
the conformally compact causal diamond it has an edge through which excitations can leak out. In
other words \emph{there are ``edges'' to the domain of dependence which break homogeneity}.
\end{itemize}

Perhaps the most important of these points is that the two sub-Hilbert spaces interact. The
associated CFTs are non-interacting only because they are defined inside causal diamonds. AdS/CFT
tells us that the dual states in the bulk should also be interacting. This is indeed the case for
the bulk $AdS_3$ geometry when viewed as two halves of the solid cylinder, even though the Rindler
wedges are oblivious to the interaction. This tells us that this particular piece of the AdS/CFT
dictionary, a couple of hyperbolic CFTs in the thermofield double state being dual to global $AdS_3$
is consistent.
 
\section{Eternal AdS}\label{sec:eternal}

Let us now move on to the case that really interests us: the eternal AdS black hole. The eternal
$AdS_d$ black hole has two asymptotic boundaries with cylindrical geometry $\re \times S^{d-2}$. The
state in the bulk is the Hartle--Hawking state~\cite{PhysRevD.13.2188}. As this geometry has two
boundaries one would expect that if there were a dual description, it would involve two decoupled
CFTs living on the two boundaries. Indeed in~\cite{Maldacena:2001kr}, Maldacena proposed that an
eternal AdS black hole is dual to two CFTs living on cylinders that are together in a thermofield
double state. The motivation for the choice of state seems to be that the bulk Hartle--Hawking
state is a thermofield double state with respect to the two outside regions~\cite{Israel:1976ur}.

This proposal has quite remarkable implications. Imagine introducing uncorrelated localised
perturbations on the two spheres $S^{d-2}$ on which the CFTs live. As before, we will call them
boundary-Alice and boundary-Bob. The duality tells us there are bulk versions of these perturbations
called bulk-Alice and bulk-Bob close to the respective boundaries. The proposed duality further
tells us that since bulk-Alice and bulk-Bob can meet behind the horizons and can, say, exchange
qubits, there must be a dual process in the two CFT-on-spheres system. Now recall that the two CFTs
are decoupled and so a meeting of such perturbations is quite counterintuitive. This proposal has
recently led to other proposals which in short can be summarised as
``entanglement=geometry''~\cite{VanRaamsdonk:2010pw,VanRaamsdonk:2013sza} and
``ER=EPR''~\cite{Maldacena:2013xja}. The idea is that if two systems are entangled they also share a
non-traversible wormhole.

Our claim is that the original proposal is incorrect and so are the ones based on it. We will go
about demonstrating this by looking at the system originally considered in~\cite{Maldacena:2001kr},
the BTZ black hole. The advantage of this system is that while it is an eternal AdS black hole with
two asymptotic boundaries in its own right, it also comes from global $AdS_3$ by orbifolding. In
fact, this orbifolding is most easily seen in the Rindler-AdS coordinates. After briefly reviewing
the construction we will show that the effect of orbifolding on the boundary is to change the causal
structure such that the process that allowed excitations to leave the causal diamond does not work
anymore. The boundary, which was two causal diamonds touching at their ends, becomes two
cylinders. It is manifest that excitations created on the two cylinders cannot meet since they
cannot leak out of cylinders unlike casual diamonds. However, the bulk does contain horizons and
excitations in the two asymptotic regions that can meet behind the horizon. We take this as evidence
that the proposal in~\cite{Maldacena:2001kr} does not work, i.e. the eternal AdS black hole is not
dual to the two CFTs on cylinders.

\subsection{BTZ as orbifold of $AdS_3$}

\begin{figure}[htbp]
\begin{center}
\subfigure[Orbifolding the bulk]{
\includegraphics[scale=.2]{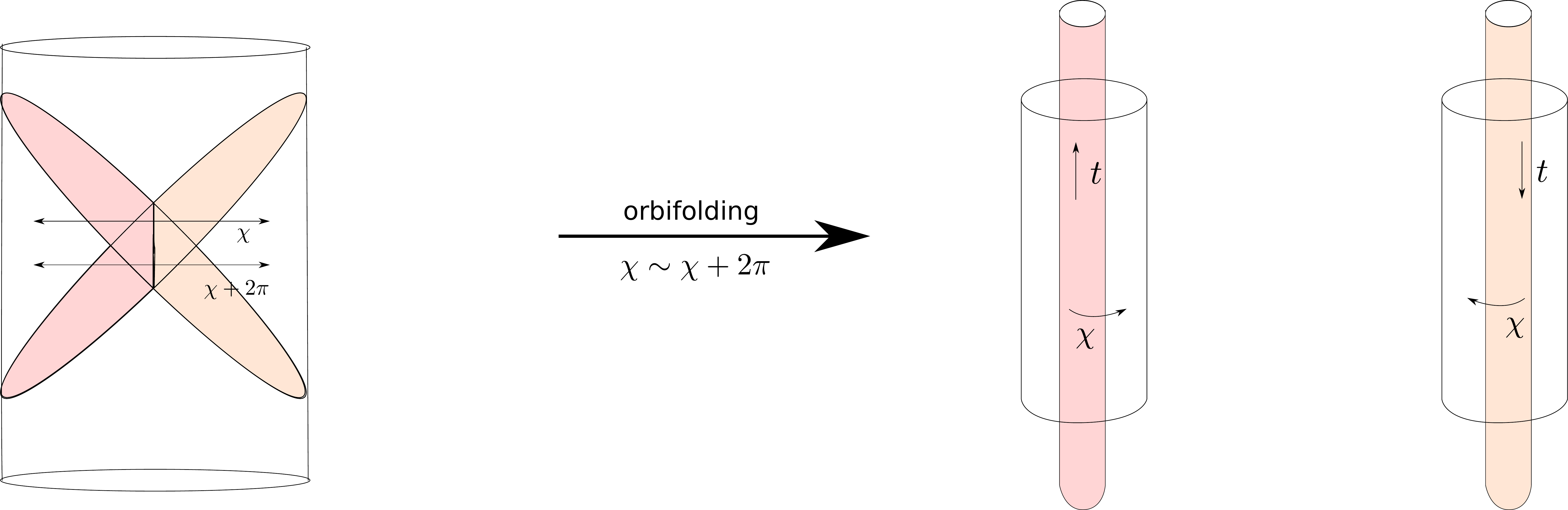} \label{OrbifoldingBulk}} \\
\subfigure[Orbifolding the boundary]{
\includegraphics[scale=.2]{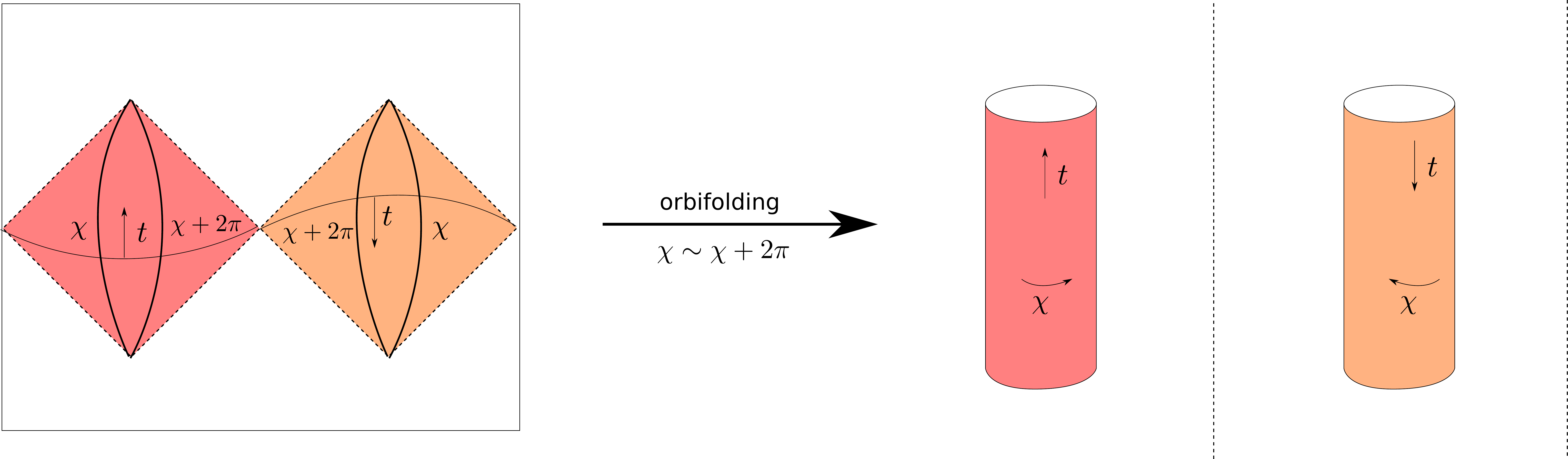} \label{OrbifoldingBoundary}} 
\caption{In (a) the effect of orbifolding in the bulk is shown. Connectivity across the horizon is maintained and thus so is the interaction. In (b) the effect of orbifolding on the boundary is shown. Hyperbolic CFTs are replaced by cylindrical ones which are not connected and do not interact.}
\label{Orbifolding}
\end{center}
\end{figure}

Following the work of~\cite{Banados:1992gq}, it is well known that the BTZ black hole can be viewed
as an orbifold of $AdS_3$. What is perhaps less well known is that said orbifolding is very
simply related to $AdS_3$ expressed in terms of Rindler-AdS coordinates. We give the details of this
in appendix~\ref{BTZasOrbifold} (see
also~\cite{Martinec:2002xq,Hemming:2002kd} for good reviews), however, for our purposes the bottom line is that the orbifolding is the identification
\be
\chi \to \chi+2 \pi \label{BTZIdentification}
\ee
on \bref{TopologicalBlackHole} defined in the Left region and its analytic continuation to the Future, Past and Right regions in Figure~\ref{RindlerAdSBulk}. The orbifolding turns the horizon into a cylinder with the periodic direction parameterised by $\chi$. The rest of the bulk outside the horizon becomes a solid annulus. Importanly, connectivity is maintained across the horizon. This is shown in Figure~\ref{OrbifoldingBulk}.

The acceleration horizon intersects the boundary cylinder at $\chi=\pm \infty$. However, under the
identification \bref{BTZIdentification} the CFT causal diamond ``becomes'' a cylinder (something we elaborate on shortly) and the bulk
horizon gets ``disconnected'' from the edge of the diamonds in the boundary. This is shown in
Figure~\ref{OrbifoldingBoundary}. This means that while we could disturb the state at the horizon by
inserting an operator in the global boundary at $\chi = \infty$ which would send a shock wave along
the relevant horizon as discussed in Section~\ref{sec:rindler} for Rindler-AdS, we cannot do so after the
orbifolding. 

\subsection{Orbifolding breaks connectivity in the boundary}

It is useful to investigate the effect of the orbifolding \bref{BTZIdentification} on the CFT in
detail as this is a potentially confusing issue. 

\begin{itemize}
\item {\bf No interaction:} Before orbifolding the two halves of the boundary cylinder were
  interacting. While this interaction was not visible in the hyperbolic CFTs, it was nevertheless
  present in the full theory and was responsible for excitations in the two CFTs meeting. Upon
  orbifolding, the two hyperbolic CFTs get replaced by two CFTs on cylinders. While non-interacting
  hyperbolic CFTs are defined on open intervals may be a part of a bigger system that may allow
  interaction, cylinders are periodic and therefore complete systems. This is clear from the
  Carter--Penrose diagrams of the two systems in Figure~\ref{OrbifoldingBoundary}.  It seems obvious
  that cylinders cannot be embedded in a bigger theory that allows interaction and therefore
  connectivity. We will nevertheless give some more evidence for this.

\item{\bf CFTs are closed systems without quasi-normal modes:} The two-point function on the
  hyperbolic CFT~\eqref{eq:twopt-hyp} is basically the two-point function of a $1+1$ dimensional CFT
  with infinite spatial extent and temperature $\frac{1}{2\pi}$ and is thus fixed by conformal
  invariance.  Naively, one would think that the two-point function after orbifolding can be
  obtained by the method of images (see~\cite{Birmingham:2002ph} for example) but this is incorrect
  for reasons we explain in Appendix~\ref{sec:Images}. One would like to know what the correct
  correlation function is to understand the properties of the bulk microstates. Unfortunately, the
  answer is not universal and will be hard to evaluate for a strongly coupled theory.\footnote{For
    correlation functions at the ``orbifold-point'' in the D1-D5 system for ``long strings'' see for
    instance~\cite{Maldacena:2001kr,Balasubramanian:2005qu}.} We do know it will be spatially
  periodic with period $2 \pi$ and will have a temporal periodicity also with a very long period
  coming from discreteness of spectra and the resulting Poincare recurrence. This rules out
  quasi-normal modes~\cite{Maldacena:2001kr,Barbon:2004ce}.  Recall, from Section
  \ref{sec:Connectivity} that a lack of mass gap and presence of quasi-normal modes in the CFT were
  necessary conditions for connectivity in the Rindler-AdS case. This is extra evidence that
  connectivity is broken when we orbifold and the two cylindrical CFTs cannot be embedded in a
  bigger theory, unlike the hyperbolic CFTs.

\item{\bf No analytic continuation:} Upon orbifolding, the CFTs' hyperbolic spacetime is replaced by
  cylinders. The analogues of the Future and Past regions of Figure~\ref{RindlerAdSBoundary} do not
  exist after orbifolding. Thus, on physical grounds it seems that boundary-Alice and boundary-Bob
  cannot meet.  However, one may still wonder if it is possible to analytically continue the correct
  non-decaying two-point function from within the fundamental region of orbifolding inside the
  diamond to outside the diamond under the map~\eqref{CFTmap} (see
  Figure~\ref{BoundaryIdentification}).
\begin{figure}[htbp]
\begin{center}
\includegraphics[scale=.5]{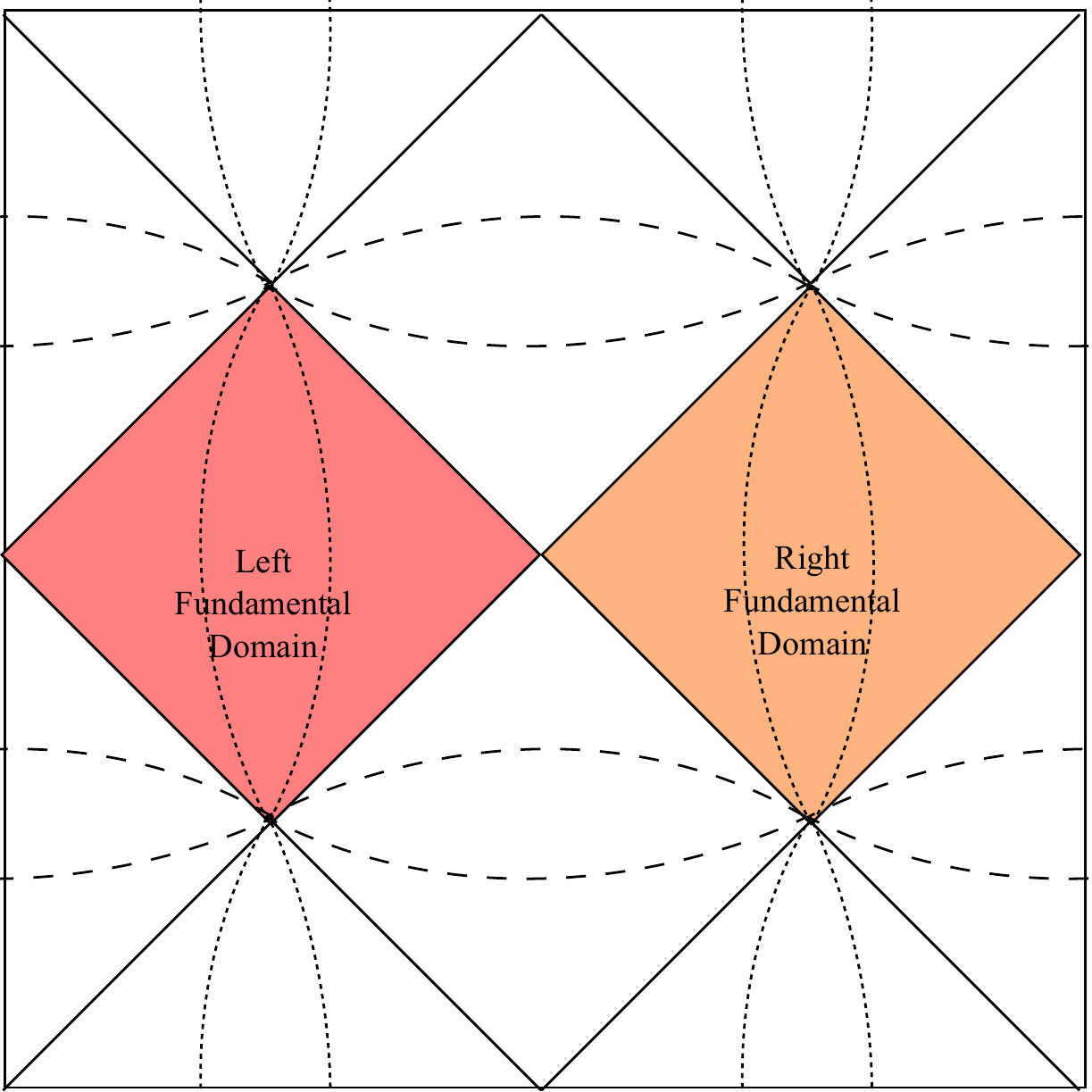}
\caption{The fundamental domains of orbifolding are obtained by identifying the dotted lines. This results in the boundary of BTZ becoming cylinders. There are also the other fundamental domains obtained by identifying the dashed lines. These produce closed timelike curves and are truncated to get the BTZ geometry. These are not relevant for our purposes. If one tries to analytically continue periodic two-point functions defined in the fundamental domain, one encounters essential singularities at the edges of the diamond. This shows that boundary-Alice and boundary Bob cannot meet in the Future region anymore. This is of course consistent with the physical intuition that cylinders, unlike causal diamonds, are not open and therefore not ``leaky'' and orbifolding thus effectively removes the Future region.}
\label{BoundaryIdentification}
\end{center}
\end{figure}
 If it were indeed possible, this
propagator could allow boundary-Alice and boundary-Bob to meet. This is of course seems nonsensical from a physical point of view but if the answer came out in the affirmative we would be forced to think that two CFTs on cylinders could somehow be embedded in a bigger theory despite the arguments given above. It turns out that reasonable periodic functions oscillate increasingly fast as one
approaches the boundary of the diamond and there is an essential singularity at the
boundary. Formally, one can still step into the complex plane and analytically continue the two
point function into the inside of the next diamond but this needs to be interpreted carefully. The
edge of the causal diamond was associated with the horizon in the bulk. We already saw that the
orbifolding disconnected the horizon from the boundary cylinder. The essential singularity at the
old connecting surface further shows that a perturbation cannot ``reach'' the edge of the diamond;
however, one can still insert operators inside two different diamonds and get finite two-point
functions. This is possibly related to matching of thermofield double correlators with geodesics
passing through the forward wedge~\cite{Kraus:2002iv,Fidkowski:2003nf}. That aside, let us emphasize that our main point is that there is no process in the CFT that captures the crossing of the horizons in the bulk.

\end{itemize}

\subsection{What does the CFT tell us about the bulk?}
 
The orbifolding that gives BTZ black holes form global $AdS_3$ replaces the non-interacting boundary
hyperbolic CFTs by non-interacting cylindrical CFTs. On physical grounds two non-interacting
cylindrical CFTs cannot be a part of a bigger system where the associated Hilbert spaces interact
(unlike two non-interacting hyperbolic CFTs). Nevertheless, we gave a number of evidences for this
fact.

Now let us understand the implication of this for the bulk. The result of Strominger--Vafa for BPS
D1-D5-P system~\cite{Strominger:1996sh}, and follow-up results for near-extremal D1-D5-P
system~\cite{Horowitz:1996fn} tell us that the dimensionality of the Hilbert space of two CFT system
is the same as that of the bulk BTZ black hole. However, as emphasised in the introduction, all
Hilbert spaces of the same dimension are isomorphic, so a particular ÒstateÓ is only meaningful if
we agree what operators we are going to act with before hand. In particular we have to specify which
Hamiltonian to evolve it with. In Section~\ref{FieldTheoryTwain} we demonstrated that the same
Minkwoski vacuum state evolves differently for the system with a mirror and one without. In
particular the system with a mirror has non-interacting left and right parts. Since, the two
cylindrical CFTs do not interact in any way, it stands to reason that their bulk duals do not
interact either. Without an interaction, they cannot have shared Future and Past
wedges. Thus, the bulk dual is manifestly not the BTZ black hole which has an interaction between the
left and right side and thus has the Future and Past wedges.

\section{Conclusions}

\subsection{Summary}

If we divide $1+1$ dimensional Minkwoski spacetime into two parts $x>0$ and $x<0$ then it seems
intuitive that moving across the shared boundary involves an interaction between the two
subsystems. Nevertheless, we demonstrated this explicitly. We also showed that
the effect of removing this interaction from the Minkowski vacuum, i.e. putting a two sided mirror
at $x=0$, is to have divergent stress tensor in the future wedge. Furthermore, after putting in the
mirror, excitations on either side cannot meet each other and hit a ``firewall'' while crossing the
Rindler horizon. Within the left and right Rindler wedges, however, the physics is indistinguishable
between the two systems. Thus Rindlerization is oblivious to the potential existence of a mirror
inserted at $x=0$.

The point of discussing this toy model is to show that connectivity across a horizon and the
possibility of Alice from the left Rindler wedge and Bob from the right Rindler wedge meeting in the
future exists only if the Hilbert spaces corresponding to the two Rindler systems interact, even
though this interaction is invisible with the Rindler evolution. Within holography, since connectivity
in the bulk comes with interactions between the two subsystems, the holographic dual field theories
also have to be interacting for the bulk to be connected.

We then discussed an example with interactions in the bulk and also in the boundary. Global AdS can
be Rindlerized and bulk-Alice and bulk-Bob introduced in Left and Right Rindler wedges can meet in
the Future region as shown in Figure~\ref{RindlerAdSBulk}. The CFTs dual to the Rindler wedges are
hyperbolic CFTs. These are defined on open intervals and are part of the global CFT. Boundary-Alice
and boundary-Bob can meet because of the interaction between the Hilbert spaces corresponding to the
two hyperbolic CFTs (although the CFTs themselves are oblivious to such interactions)  as shown in
Figure~\ref{RindlerAdSBoundary}. Thus the bulk duals to the hyperbolic CFT should be connected and
this is consistent with the bulk dual being global AdS.
 
Finally, we turned to the eponymous subject of the paper, the eternal AdS black hole and its
purported dual non-interacting thermofield double CFT system on cylinders.  We specifically
discussed the BTZ black hole, which has the advantage of being obtainable by orbifolding Rindler-AdS;
this allows easy comparison between Rindler-AdS and the BTZ black hole. The effect of orbifolding in
the bulk is to replace the hyperbolic horizon by a cylindrical one. This maintains the smoothness
across the horizon and shows that the two Hilbert spaces in the bulk remained interacting after the
orbifolding. On the other hand, the effect of orbifolding on the boundary is radically different. It
replaces hyperbolic CFTs by cylindrical CFTs. While it is obvious the two CFTs cannot be embedded
in a bigger system to make the associated Hilbert spaces interacting, we nevertheless gave further
arguments to show this. This lack of interaction in the boundary implies that the bulk dual systems
also do not interact. This rules out the possibility of the BTZ being the bulk dual.
 
\subsection{Discussion}
 
Having established that the thermofield double state of two CFTs on cylinders is not dual to the
eternal AdS black hole, what is the dual bulk description? No connectivity between the two systems
in the dual picture is like the situation with the mirror in the toy example of
Section~\ref{FieldTheoryTwain}. If the analogy were exact then the bulk states would be identical
to the black hole all the way to the horizon and abruptly change. However, one does not expect to be
able to make a mirror of infinitesimal thickness in quantum gravity and a reasonable guess would be
that the mirror width is Planck scale or string scale. Thus, it seems that the dual geometry would
resemble a black hole outside the stretched horizon on both sides, but due to an absence of interactions
between the two stretched horizons there would be no future and past wedges.
 
\begin{figure}
\begin{center}
\subfigure[``Entanglement=Spacetime'']{\label{fig:vanRaamsdonkPicture}\includegraphics[scale=.22]{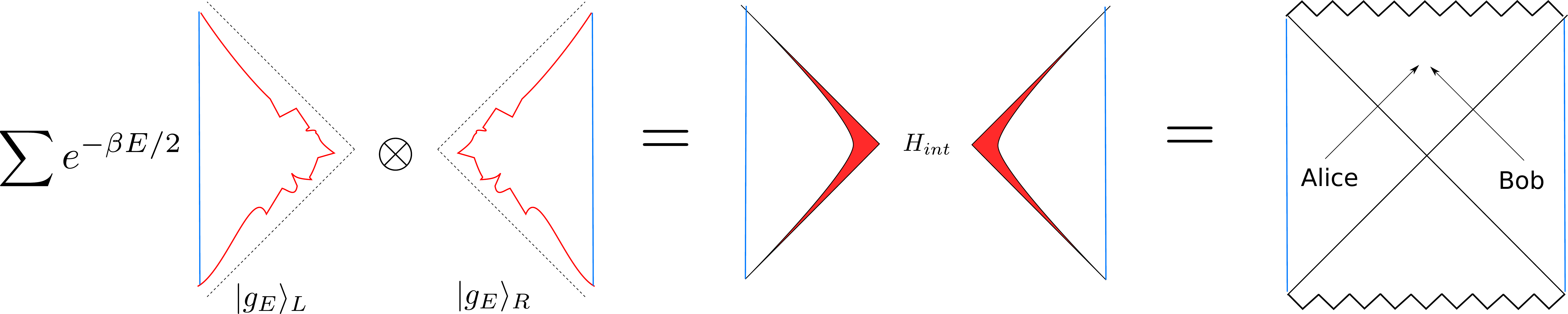}} \\
\subfigure[``Entanglement is not enough'']{\label{fig:OurPicture}\includegraphics[scale=.22]{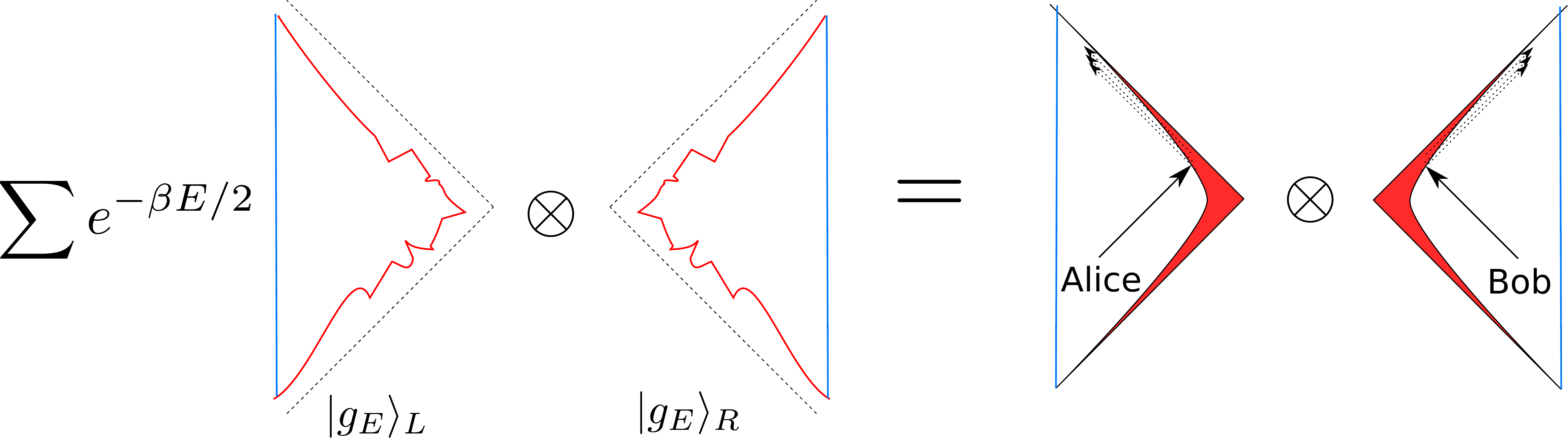}
}
\end{center}
\caption{ The idea advocated in~\cite{Czech:2012be}, based on the proposal of~\cite{Maldacena:2001kr}, (a) is that generic geometric-microstates dual to decoupled CFTs resemble a black hole outside the stretched horizon but then differ sharply. However, the thermofield double state would involve a topology change in the bulk which creates an eternal AdS black hole. Alice and Bob falling from different sides can then meet in the future wedge. In (b) we summarise our results which is that for all kinds of generic states including the thermofield double state there is no forward wedge. The wedges end in fuzzballs, which maybe thought of as a microscopic realisation of the stretched horizon. Alice and Bob falling on the two sides thermalise on the fuzz but do not meet each other.} \label{fig:TwoPictures}
\end{figure}

This picture is deceptively close to the one advocated in~\cite{Czech:2012be} where the bulk duals
to generic states of decoupled CFTs are claimed to be similar to black holes up to the stretched
horizon, but then differ sharply and have no future and past regions. However, the claim there is
that if the bulk duals are in the thermofield double state then there is a topology change in the
bulk leading to future and past wedges. Our claim is that even if the two sides are in a thermofield
double state, there are no future or past wedges. These two viewpoints are summarised in
Figure~\ref{fig:TwoPictures}.
While at this level of analysis, we cannot say much about the nature of the ``non-completable''
stretched horizons in Figure~\ref{fig:OurPicture} except that there is no interior description of a
common shared future wedge, we expect the UV complete description of the geometry to replace the
stretched horizon by fuzzball microstates. Alice and Bob falling into either side will thermalise
after hitting the respective ``fuzz'' and in time be re-emitted in a unitary fashion as shown in
Figure~\ref{fig:OurPicture}.

Our analysis has been for a particular eternal AdS black hole, the BTZ black hole, which is specific
to three spacetime dimensions. This made some arguments clearer since we could then contrast it with
the case of Rindler-AdS as the two are related by orbifolding. However, the lessons are quite
general. The two decoupled CFTs on $S^d \times \re$ do not exhibit decaying correlators. Thus, we do
not expect them to be embeddable in a bigger system and have their Hilbert spaces interact. If they
cannot be made interacting then their bulk duals would also be disconnected, and we expect a picture
like that shown in Figure~\ref{fig:OurPicture}.

We should mention that we used subregion duality crucially in the above
discussion. Ref.~\cite{Papadodimas:2012aq} has an alternate proposal where the outside of the
horizon is captured by a coarse-grained subsector of the CFT and the inside by a fine-grained
one. First, let us emphasise that this does not affect our main result that two CFTs in the
thermofield double state do not have a bulk description with a \emph{shared} future (and past)
wedge. Second, in the context of obtaining BTZ from orbifolding Rindler-AdS we find the proposal
incorrect. If the CFT were indeed to split into two subsectors then in the case of global AdS the
two Rindler wedges are entangled, so the two coarse-grained subsectors have to be entangled. The
fine-grained ones may or may not be depending on the proposal. However, upon orbifolding if each
side is to become a one-sided black hole (instead of fuzzballs ending on the stretched horizon), the
orbifolding has to make each side's coarse-grained subsector maximally entangled with the
fine-grained one. We do not see how an operation like orbifolding can accomplish the entanglement
swapping.

Finally, we would like to comment on the implications of our work for states dual to just one
CFT. Our main result is that when two cylindrical CFTs are in the thermofield double state, or any
other state for that matter, the bulk description does not involve shared future and past
wedges. Further, assuming a subregion duality of the kind advocated
in~\cite{Czech:2012bh,Hubeny:2012wa,Bousso:2012mh}, the CFT duals match the black hole outside the
stretched horizon and then start differing and capping off outside the semiclassically anticipated
horizon. In effect, the fact that the two CFTs were entangled made no difference. Then it is easy to
see that typical states of one CFT would have the same behaviour as the ensemble and the geometric
dual will end in a fuzz cap at the stretched horizon with there being no region behind the future
and past horizons. This is to be contrasted with the traditional picture where an excitation in a
single CFT settles down to a pure state which is described by a black hole formed from collapse.

It would of course be interesting to see how this comes about and the detailed dynamics that prevent
a black hole from forming in the bulk, despite being predicted by equations of motion. One
possibility has been advocated by Mathur~\cite{Mathur:2008kg} where the exponential number of bulk fuzzball
microstates invalidates the saddle point approximation, allowing a thin shell to tunnel into (a
superposition of) fuzzball states. This still leaves open the question of what the general
mechanism responsible for the existence of so many states with horizon-scale structure is.

If, as we have argued from AdS/CFT, there are unanticipated quantum gravity effects well below the
Planck scale (under special circumstances), this opens up the possibility of other new
effects. Obviously, we do \emph{not} expect new quantum gravitational physics on solar system or
intergalactic scales, but there could be important effects in early cosmology.
 
\section*{Acknowledgements}

We would like to thank Jan de Boer, Saugata Chatterjee, Bartek Czech, Damien Easson, Cindy Keeler,
Finn Larsen, Gilad Lifschytz, Andrew Long, David Lowe, Juan Maldacena, Emil Martinec, Samir Mathur,
Niels Obers, Maulik Parikh, Eray Sabancilar, Masaki Shigemori,  Joan Simon and Tanmay Vachaspati for
helpful discussions. We would like to thank the organisers of the ``Black holes: Complementarity,
Fuzz or Fire'' workshop at KITP, Santa Barbara for their hospitality during the early phase of this
project.  BDC would like to thank the organisers of the workshop ``Black holes in string theory
workshop'' at MCTP, Ann Arbor for their hospitality during the intermediate phase of this work. SA was supported in part by DOE grant de-sc0010010 and an FQXi grant.

\appendix

\section{Global vs. Rindler-AdS coordinates} \label{variousCood}

Often global AdS is written in terms of coordinate $\rho, \tau$ and the angular coordinates as
\be
ds^2= -(1+ \rho^2) d\tau^2 + \f{d\rho^2}{1+ \rho^2} + \rho^2 d\Omega_{d-1}
\ee
however the boundary is then at $\rho=\infty$. To see the physics of the boundary better, it is useful to use a different radial coordinate given by the relation $\tan u=\rho$. Then the metric is
\be
ds^2 = \sec^2 u ( -d\tau^2 + du^2 + \sin^2 u ~d\Omega_{d-1}^2)
\ee
where the radial coordinate has the range given by $u \in [0,\pi/2)$. The Rindler AdS-coordinates treat the polar angle of $S^{d-1}$, $\phi$, separately and are related to global coordinates by
\bea
 \tan^2 u &=&  (r^2-1) ~ \f{ \cosh (2 \chi) + \cosh (2 t) }{2} + \sinh^2 (\chi), \label{CoordRela}\\
\tan \phi &=& \f{r}{\sqrt{r^2-1}} \f{\sinh(\chi)}{ \cosh(t)},  \label{CoordRelb} \\
\tan \tau &=& \f{\sqrt{r^2-1}}{r} \f{\sinh(t)}{ \cosh ( \chi)}. \label{CoordRelc}
\eea
We note that $\phi \in [0, \pi]$ for $d>1$ and $\phi \in [0, 2\pi)$ with $\phi \sim \phi+2 \pi$ for $d=1$. The inverse relations are given by
\bea
 r^2 -1 &=&- \sin^2 \tau \sec^2 u +\cos^2 \phi \tan^2 u, \\ \label{CoordRel2a}
\coth \chi &=& \f{\cos \tau}{\sin \phi} \f{1}{\sin u}, \\ \label{CoordRel2b}
\tanh  t &=& \f{\sin \tau}{\cos \phi} \f{1}{\sin u}  \label{CoordRel2c}
\eea
where the radial coordinate is written shifted so that the RHS gives the Rindler horizon. Note that $\chi \in (-\infty,\infty)$. In terms of these coordinates the metric is
\be
ds^2 =- (r^2-1) dt^2 + \f{d r^2}{r^2-1} +r^2 \left[ d \chi^2 + \sinh^2 (\chi) d\Omega_{d-2}^2 \right] \label{TopologicalBlackHoleAppendix}. 
\ee 

From \bref{CoordRel2a}  we see that the intersection of the Rindler horizon $r=1$ and the boundary cylinder $u= \pi/2$ is given by
\be
\sin \tau = \pm \cos \phi.
\ee
From this it is easy to see that at the edges of the causal diamond $t,\chi = \pm \infty$.  We  will see that orbifolding $AdS_3$ to get BTZ requires $\chi \sim \chi+2 \pi$ and therefore we see that the horizon gets disconnected from the boundary cylinder.

\section{BTZ as an orbifold of $AdS_3$} \label{BTZasOrbifold}

It is well known that BTZ can be viewed as an orbifold of $AdS_3$~\cite{Banados:1992gq}.  We follow the analysis in \cite{Martinec:2002xq}.  $AdS_3$ is the group manifold of $SL(2,\mathbb R)$ and one can quotient it
\be
g \sim hgh
\ee
where $h$ is a hyperbolic element of $SL(2,\mathbb R)$ to obtain the BTZ geometry.
Let us parameterise $g$ as
\bea
g &=&   \left(
      \begin{array}{cc}
       a  & b  \\
     c &  d 
      \end{array} \right)  \nn
&=& \left(
      \begin{array}{cc}
        \cos(\tau)\sec(u) + \sin(\phi) \tan(u)  & \sin(\tau)\sec(u) + \cos(\phi) \tan(u)  \\
     - \sin(\tau)\sec(u) + \cos(\phi) \tan(u)  &  \cos(\tau)\sec(u) - \sin(\phi) \tan(u)  
      \end{array} \right)
\eea
From \bref{CoordRela},\bref{CoordRelb} and \bref{CoordRelc} we see that the Rindler-AdS coordinates are related to $g$ by
\begin{gather}
a=  r ~e^{ \chi}, ~~ b= \sqrt{ r^2-1}~ e^{ t} \\
c=\sqrt{ r^2-1} ~ e^{- t},~~ d= r ~ e^{- \chi}.
\end{gather}
The metric is then given by
\be
ds^2 =- (r^2-1) dt^2 + \f{d r^2}{r^2-1} +r^2  d \chi^2.
\ee
These coordinates only cover one of the exterior regions but one can analytically continue to the other exterior and the future and past regions by the Kruskal extension
\be
\mathrm u=- \sqrt{\f{r-1}{r+1}} e^{-t}, \qquad \mathrm v= \sqrt{\f{r-1}{r+1}} e^{t}.
\ee
They are related to $g$ as
 \begin{gather}
a=  \f{1-\mathrm u \mathrm v}{1+\mathrm u \mathrm v} e^{ \chi}, ~~ b= -\f{2 \mathrm u}{1+\mathrm u \mathrm v}~ e^{ t} \\
c=\f{2 \mathrm v}{1+\mathrm u \mathrm v} ~ e^{- t},~~ d= \f{1- \mathrm u \mathrm v}{1+ \mathrm u \mathrm v} ~ e^{- \chi}.
\end{gather}
The metric is then given by
\be
ds^2 = \f{1}{(1+ \mathrm  u \mathrm v)^2}(-4 d\mathrm u d\mathrm v + (1- \mathrm u \mathrm v)^2 d \chi^2).
\ee
If we quotient  $AdS_3$ by  $h=e^{\pi \sigma_3}$ we get
\bea
&&  \left(
      \begin{array}{cc}
        \cos(\tau)\sec(u) + \sin(\phi) \tan(u)  & \sin(\tau)\sec(u) + \cos(\phi) \tan(u)  \\
     - \sin(\tau)\sec(u) + \cos(\phi) \tan(u)  &  \cos(\tau)\sec(u) - \sin(\phi) \tan(u)  
      \end{array} \right)  \nn
 &=&\left(
      \begin{array}{cc}
       e^{2\pi}( \cos(\tau)\sec(u) + \sin(\phi) \tan(u))  & \sin(\tau)\sec(u) + \cos(\phi) \tan(u)  \\
     - \sin(\tau)\sec(u) + \cos(\phi) \tan(u)  &  e^{-2 \pi}(\cos(\tau)\sec(u) - \sin(\phi) \tan(u)  )
      \end{array} \right) 
\eea
In particular this implies 
\be
\chi \sim \chi+2 \pi
\ee
 which converts the Rindler-AdS geometry which has the same metric as the BTZ black hole, into a true BTZ black hole with a cylindrical boundary. The BTZ horizons are at $\mathrm u \mathrm v=0$, and the boundary at $\mathrm u \mathrm v=-1$. There is also a orbifold singularity at $\mathrm u \mathrm v=1$ beyond which the geometry is cutoff because of closed timelike curves.
 
\section{Infall as Bell measurement}\label{sec:bell}

Immediately after AMPS's firewall paper~\cite{Almheiri:2012rt}, one of us proposed that the correct
way to analyse the situation would be think of the observer as part of the complete system and
measurements as coming from decoherence between the observer (or her apparatus) and the rest of the
system~\cite{Chowdhury:2012vd}. This appendix is based on the same theme and can be seen as a
motivation for the rest of the paper. We will consider bulk-Alice, a closed string object moving
around in the eternal AdS black hole, and ask what does it take for her to verify that the horizon
is smooth. We will then consider boundary-Alice, an open string excitation moving on one of the
boundary cylinders, and ask the same question. We will demonstrate an inconsistency and claim that
the thermofield double CFTs are not dual to the eternal AdS black hole.

Consider massless fields in $1+1$ dimensions. The equations of motion split the fields into left and
right movers. We consider only the left movers and the right movers behave the same. It can shown
(see~\cite{Parentani:1993yz} for example) that the Minkowski vacuum can be expressed in terms of
Rindler modes as
\be
\ket{0_M} = \f{1}{\sqrt{\prod Z_\lambda}} \prod_\lambda e^{\tanh \theta_\lambda b_{\lambda,R}^\dagger b_{\lambda,L}^\dagger}\ket{0_R}\ket{0_L}.
\ee
where $Z_\lambda = Tr[e^{-2 \pi \lambda/a}]$ and  $\tanh \theta_\lambda =e^{-\pi \lambda/a}$ where $a$ is the acceleration of the Rindler observer. Different modes given by different $\lambda$ decouple and we can focus on the vacuum for a particular $\lambda$ 
\be
\ket{0_{M,\lambda}} = \f{1}{\sqrt{Z_\lambda}} \sum \tanh^n \theta_\lambda~ \ket{n_{\lambda,R}} \ket{n_{\lambda,L}}.
\ee
Note that if we consider the high temperature limit and restrict to fermionic modes then the above truncates to
\be
\ket{0_{M,\lambda}}  = \f{1}{\sqrt{2}} ( \ket{0_{\lambda,R}} \ket{0_{\lambda,L}} +  \ket{1_{\lambda,R}} \ket{1_{\lambda,L}} ) \label{MinkowskiMode}
\ee
and we can simplify our analysis by just considering qubits. The right moving observer will encounter left moving modes localised inside and outside the horizon and will find the state as the vacuum only if together they are in the state \bref{MinkowskiMode}. 

There is a simple generalization of the Minkwoski vacuum state~\bref{MinkowskiMode} which is a maximally entangled state between the two subsystems. One can write down four such orthogonal states
\begin{equation}\begin{aligned}
\ket{\varphi_1} &\defeq \frac{1}{\sqrt{2}}\big(\ket{\hat{0} }\ket{0 } 
                           + \ket{\hat{1} }\ket{1 }\big)\,,\\
\ket{\varphi_2} &\defeq \frac{1}{\sqrt{2}}\big(\ket{\hat{0} }\ket{0 } 
                           - \ket{\hat{1} }\ket{1 }\big)\,,\\
\ket{\varphi_3} &\defeq \frac{1}{\sqrt{2}}\big(\ket{\hat{0} }\ket{1 } 
                           + \ket{\hat{1} }\ket{0 }\big)\,,\\
\ket{\varphi_4} &\defeq \frac{1}{\sqrt{2}}\big(\ket{\hat{0} }\ket{1 } 
                           - \ket{\hat{1} }\ket{0 }\big)\,, \\ \label{BellStates}
\end{aligned}\end{equation}
and these are referred to as Bell states. The $\ket{\hat 0}$ and $\ket{\hat 1}$ are eigenstates of $\hat \sigma_z$ and similarly $\ket{0}$ and $\ket{1}$ are eigenstates of $\sigma_z$.Observe that in a simplified qubit model the Minkowski state corresponds to the first Bell state. 

The reduced density matrix of the hatted and the  unhatted systems for all four states are
\be
\hat \rho  = \h ( \ket{\hat 0} \bra{\hat 0} +  \ket{\hat 1} \bra{\hat 1} ), \qquad \rho  = \h ( \ket{0} \bra{0} +  \ket{1} \bra{1} )
\ee
which means that Charlie with access to only one of the systems (i.e. with access to operators $\hat I \otimes \sigma_x,~\hat I \otimes \sigma_y$ and $\hat I \otimes \sigma_z$) will get identical response from all four states and will be unable to distinguish between them. This does not, however, mean that the four states are indistinguishable. These states are eigenstates of the operators $\hat \sigma_x \otimes \sigma_x,~\hat \sigma_y \otimes \sigma_y$ and $\hat \sigma_z \otimes \sigma_z$. The eignevalues are shown in the table below.
\begin{center}
\begin{tabular}{|c|c|c|c|c|}
\hline
\text{state} & $\hat \sigma_x \otimes \sigma_x$ & $\hat \sigma_y \otimes \sigma_y$ & $\hat \sigma_z \otimes \sigma_z$ \\
\hline
\hline
$\ket{\varphi_1}$ & +1 & -1 & +1 \\
$\ket{\varphi_2}$ & -1 & +1 & +1 \\
$\ket{\varphi_3}$ & +1 & +1 & -1 \\
$\ket{\varphi_4}$ & -1 & -1 & -1 \\
\hline
\end{tabular} 
\end{center}
Thus, an observer, Alice, can distinguish between the four states by measuring the expectation value of any of the two operators, say $\hat \sigma_x \otimes \sigma_x$ and $\hat \sigma_z \otimes \sigma_z$. This is called a Bell measurement. 

\begin{figure}[htbp]
\begin{center}
\subfigure[Non-Bell Measurement]{
\includegraphics[scale=.35]{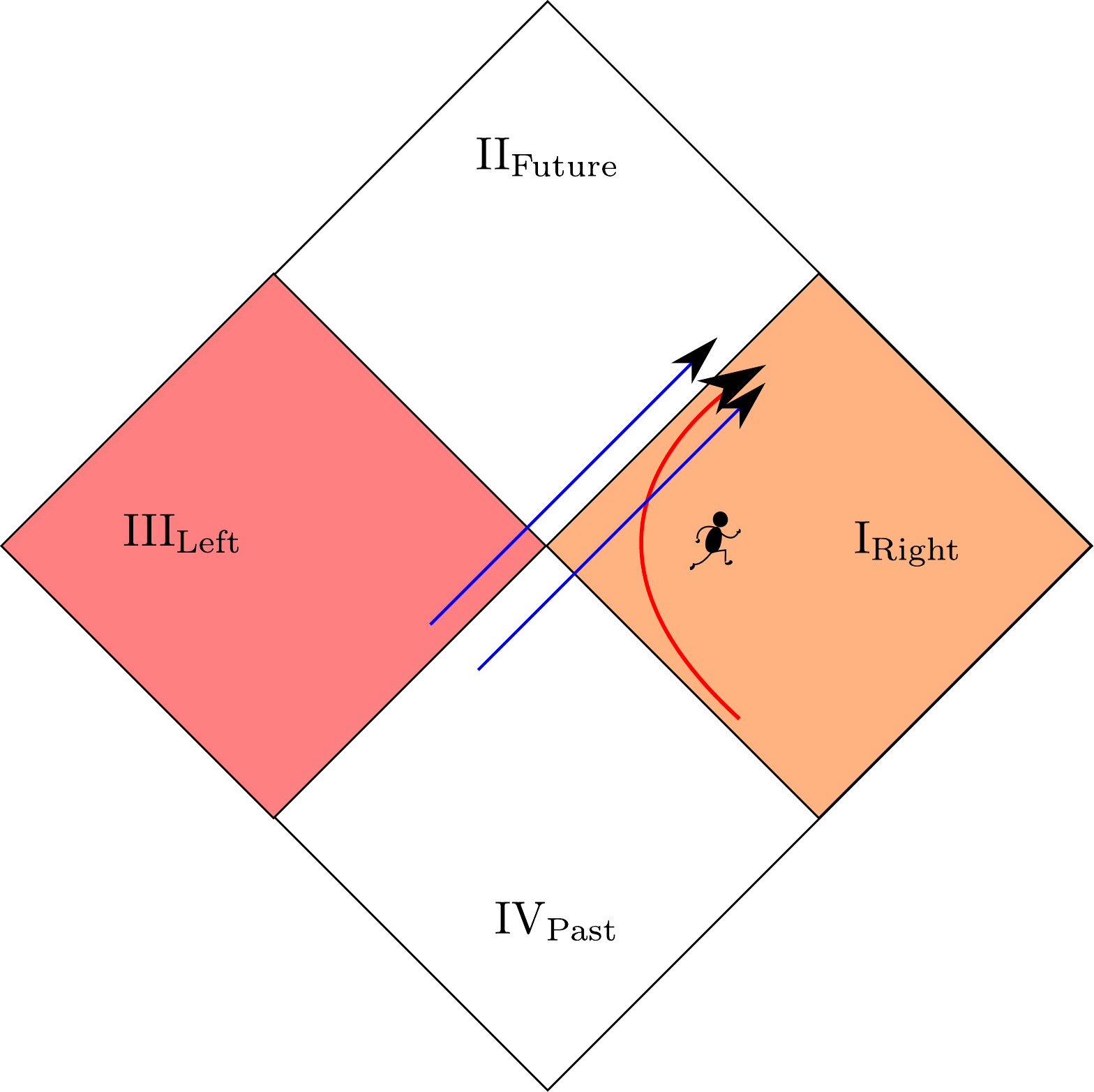} \label{NonBell}} \hspace{2cm}
\subfigure[Bell Measurement]{
\includegraphics[scale=.35]{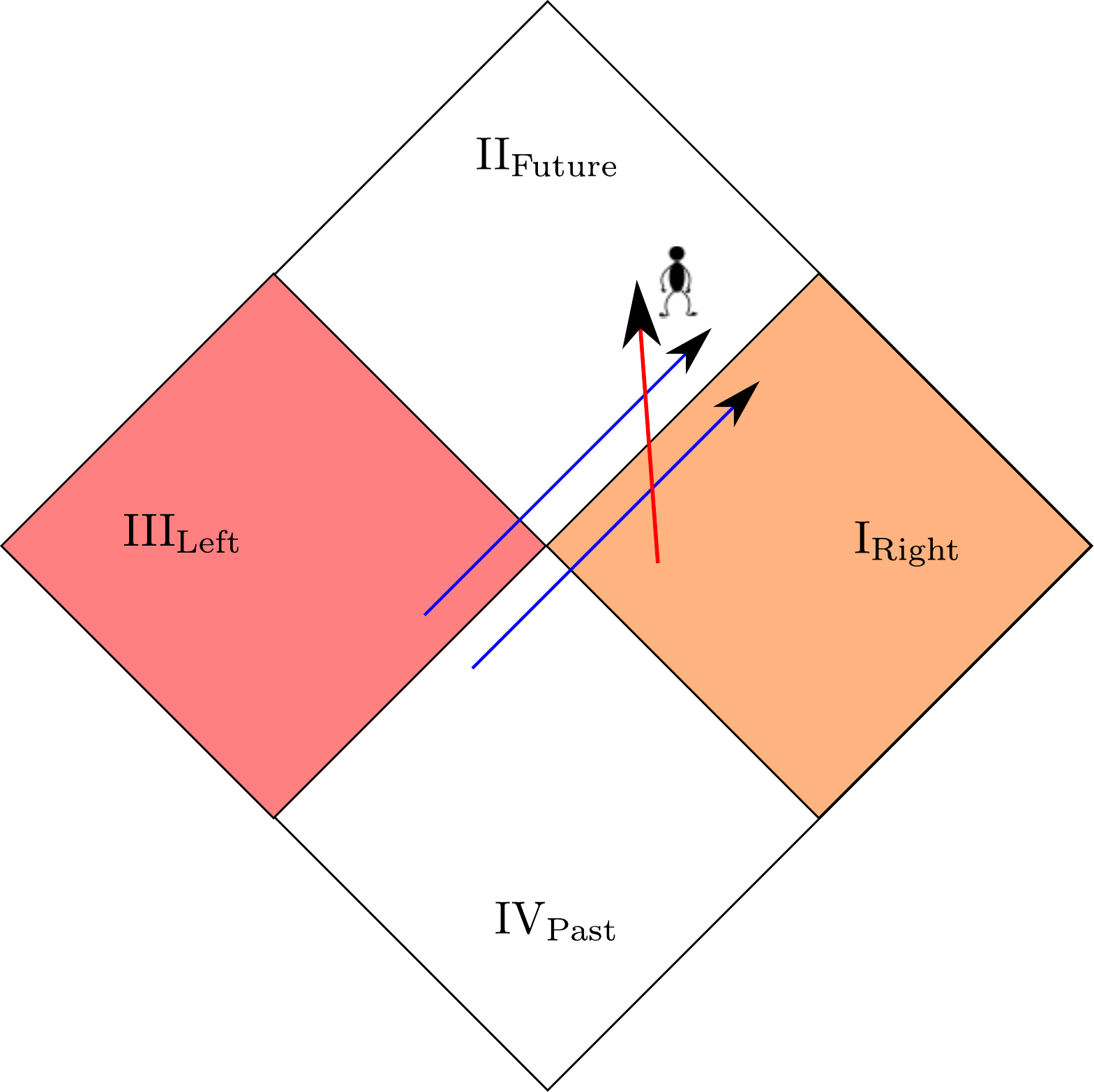} \label{Bell}} 
\caption{An accelerating observer only intersects the modes of the right wedge so can only do non-Bell measurements. These do not measure the actual state of the system but instead collapse the system into a different state. An inertial observer on the other hand intersects both modes and thus can perform a Bell measurement to verify that the full state is the Minkowski vacuum. }
\label{RindlerMinkowski}
\end{center}
\end{figure}

In light of this, our previous comment about a right moving observer finding the left movers in the vacuum only if they are in the state \bref{MinkowskiMode} can be restated in the following way. Accelerating observers who stay inside the Rindler wedge have access to only half the system can only perform non-Bell measurements and cannot tell of the full state is the Minkwoski vacuum or any other state that leaves the right wedge density matrix the same (see Figure~\ref{NonBell}). However, inertial observers can measure the full state of the system and in fact do so while crossing the horizon. They can thus tell if the full state is the Minkwoski vacuum or some other state. Thus inertial observers perform Bell measurements. This is shown in Figure~\ref{Bell}.

This has immediate consequences for Maldacena's proposal that the eternal AdS black hole is dual to
two decoupled CFTs on cylinders in a thermofield double state~\cite{Maldacena:2001kr}. Bulk-Alice, a
creature created in the left exterior of the eternal AdS black hole (see Figure~\ref{fig:vanRaamsdonkPicture}) performs a
Bell measurement on the modes inside and outside while falling into the black hole. According to the
AdS/CFT proposal, when bulk-Alice is close to the corresponding boundary, the dual boundary-Alice is
an excitation in the CFT living on that boundary. However, since the two CFTs are defined on
cylinders and there is no interaction term between them, there is no way boundary-Alice can perform
a Bell measurement to ascertain that the two CFTs are indeed in the thermofield double state. Thus,
there is a contradiction. Since, there is nothing wrong with the bulk or the boundary systems by
themselves, the conclusion is that they are not dual to each other.
  
 \section{Why the method of images gives the wrong correlation function} \label{sec:Images}

 Naively one would think that the correlator after orbifolding the hyperbolic CFT can be obtained by the method of
images from \bref{eq:twopt-hyp} 
\be
\langle O(t,\chi) O(0,0) \rangle_{orb}^{\text{naive}} = \sum_{n=-\infty}^\infty \f{1}{16}~ \f{1}{ \sinh^2( \f{ t-\chi-2 n \pi}{2}) \sinh^2( \f{ t+\chi+2 n \pi}{2})}. \label{eq:twopt-hyp-images}
\ee
However, this is incorrect. The point is that~\bref{eq:twopt-hyp} decays with
time and thus when Fourier transformed has quasi-normal modes i.e. it has poles away from the
real line~\cite{Birmingham:2001pj,Birmingham:2002ph}. This comes from the CFT being defined on an infinite line and
physically corresponds to a wave function exploring the infinite phase space.  However, on closed
intervals the phase space is finite and while there is quasi-periodicity, there are no quasi-normal
modes~\cite{Maldacena:2001kr,Barbon:2004ce}. The problem with the correlation function obtained from
method of images \bref{eq:twopt-hyp-images} is that it still has quasi-normal modes so it cannot be the correct correlation function for a CFT of finite extent and temperature.

The problem is easily illustrated even at the level of free fields. Consider a
circular string whose transverse oscillations are governed by a CFT (it could be the closed string
of string theory). Then it has left movers and right movers and we have
\begin{equation}\begin{aligned}
\partial X &= \sum_{n \in \ints^+} \sqrt{\f{\omega_n}{2 L}} (a_{L,\omega_n} e^{-i \omega_n x_-} +a_{L,\omega_n}^\dagger e^{i \omega_n x_-}  ) \\
\bar \partial X &= \sum_{n \in \ints^+} \sqrt{\f{\omega_n}{2L}} (a_{R,\omega_n} e^{-i \omega_n x_+} +a_{R,\omega_n}^\dagger e^{i \omega_n x_-}  ) 
\end{aligned}\end{equation}
where $\omega_n = \f{2 \pi n}{L}$ and $x_\pm = t \pm y$. For a thermal state with inverse temperatures $\beta_L, \beta_R$ we get the correlator
\be
 \langle (\partial X \bar \partial X)(t,y) (\partial X \bar \partial X)(0) \rangle_{\beta_L, \beta_R} =\sum_{n \in\ints} \f{\omega_n}{2L} \f{1}{e^{\beta_L \omega_n}-1} e^{i \omega_n x_-}  \times \sum_{n \in\ints} \f{\omega_n}{2L} \f{1}{e^{\beta_R \omega_n}-1} e^{i \omega_n x_+}. 
\ee
If we approximate the summation by an integral we get
\be
\sum_{n \in\ints} \f{\omega_n}{2L} \f{1}{e^{\beta_L \omega_n}-1} e^{i \omega_n x_-}  \to \f{\pi}{4 \beta^2} \f{1}{\sinh(\f{\pi x_-}{\beta})^2}
\ee
Which is the same as \bref{eq:twopt-hyp} upto a conventional normalisation factor. This obviously had to work because the two-point function is fixed by conformal invariance when we
take $L \to \infty$, which is basically what we do when we replace the sum by an integral. However,
this example clearly shows what happens. The two-point function with the sum did not have any
quasi-normal modes and was the correct two-point function on the cylinder. If we first take an
infinite cylinder limit and then use the method of images we obviously do not recover the correct
answer.

For a CFT at finite temperature and with compact spatial extent (thus on a torus upon
Euclideanization) the two-point function is not universal but the method of images gives the {\em wrong} universal answer. 

\bibliographystyle{toine}
\bibliography{Papers}

\end{document}